\documentclass[groupedaddress,showkeys]{revtex4-2}
\usepackage{float}
\usepackage{multirow}
\usepackage{amsmath}
\usepackage{amssymb}
\usepackage{graphicx}
\usepackage{xcolor}
\usepackage{stackrel}
\usepackage{natbib}
\definecolor{tl}{RGB}{0,180,120}
\usepackage[colorlinks,citecolor=tl,linkcolor=red,urlcolor=magenta,unicode=true]{hyperref}

\begin{document}
	\title{Global 21-cm brightness temperature in  viscous dark energy models}
	
	\author{Ashadul Halder}
	\email{ashadul.halder@gmail.com}
	\affiliation{S. N. Bose National Centre for Basic Sciences,\\ JD Block, Sector III, Salt lake city, Kolkata-700106, India.}
	
	\author{Shashank Shekhar Pandey}
	\email{shashankshekhar.pandey@bose.res.in}
	\affiliation{S. N. Bose National Centre for Basic Sciences,\\ JD Block, Sector III, Salt lake city, Kolkata-700106, India.}
	
	\author{A. S. Majumdar}
	\email{archan@bose.res.in}
	\affiliation{S. N. Bose National Centre for Basic Sciences,\\ JD Block, Sector III, Salt lake city, Kolkata-700106, India.}
	
	\begin{abstract}
		\begin{center}
			\large{\bf Abstract}
		\end{center}
	
We investigate the global 21-cm brightness temperature  in the context of viscous dark energy (VDE) models. The bulk viscosity of dark energy perturbs the Hubble evolution of the Universe which could cool baryons faster, and hence, alter the 21-cm brightness temperature. An additional amount of entropy is also produced as an outcome of the viscous flow. We study the combined contribution of Hawking radiation from primordial black holes, decay and annihilation of particle dark matter and baryon-dark matter scattering in the backdrop of VDE models towards modification of the 21-cm temperature. We obtain bounds on the VDE model parameters which can account for the observational excess of the  EDGES experiment ($-500^{+200}_{-500}$ mK at redshift $14<z<20$)	 due to the interplay of the above effects. Moreover, our analysis yields modified constraints on the dark matter mass and scattering cross-section compared to the case of the $\Lambda$CDM model.

	\end{abstract}
\maketitle
	

	\section{\label{sec:intro} Introduction}	
	
	The redshifted signature of the 21-cm hydrogen absorption spectra \cite{1951Natur.168..356E,1951Natur.168..357M} has turned out to be a promising probe in exploring several unknown mysteries of the early Universe, in particular during the cosmic dark age \cite{Pritchard_2012,Chowdhury_2020}. The 21-cm line is originated as an outcome of hyperfine transition between energy levels of the neutral hydrogen atom. According to the standard $\Lambda$CDM model, the brightness temperature is  $\approx-200$ mK at Redshift $(z)\approx17$. This particular feature has come under close scrutiny due to a
	couple of recent observational results. The ``Experiment to Detect the Global Epoch of Reionization Signature'' (EDGES) \cite{edges}, 
	reports an observational excess of $-500^{+200}_{-500}$ mK at $14<z<20$. On the
	other, another recent observation SARAS \cite{saras} is at variance with the EDGES's results.  However, a simple $\Lambda$CDM analysis fails to take into
	account additional thermal contributions on the brightness temperature of the 21-cm line, such as due to the decay of the constituents of 
	dark matter, and the interaction of baryons with
	background photons.
	
	The nature of the dark sector components, {\it viz.},  dark matter and dark energy of the Universe are still unknown. In certain models baryon-dark matter interaction could lead to significant cooling of the baryonic fluid \cite{munoz,21cm_upala,rupadi,21cm_feb,PhysRevD.98.023013,PhysRevLett.121.011102,PhysRevD.98.103005}. Other phenomena that impact the baryon temperature include decay/annihilation of normal dark matter candidates \cite{rupadi,BH_21cm_1,PhysRevLett.121.011102,Mitridate_2018,PhysRevD.98.023501} and/or super-heavy dark matter \cite{21cm_mar}, dark matter - dark energy interaction \cite{upala,21cm_mar,li_IDE,idem0,idem1,idem2,idem3,Kumar_2017,Kumar_2019,PhysRevD.101.063502,DIVALENTINO2020100666}, viscosity of the dark matter fluid \cite{PhysRevD.100.063539} and other forms of energy injection in the pre-recombination era \cite{Brahma:2020tmk}. Primordial black holes may constitute a significant fraction of dark matter \cite{Carr_2016,Garc_a_Bellido_2017,Carr_2021} and their evaporation impacts the temperature of the 21-cm signal, as well \cite{BH_21cm_0,BH_21cm_1,BH_21cm_2,BH_21cm_4,BH_21cm_5,PhysRevD.98.023503,10.1093/mnras/stt1493,21cm_feb,Auffinger:2022khh,Mittal_2022,yy1,yy2}.
	
	Alongside dark matter, the currently dominating part of the Universe, {\it i.e.}, dark energy (DE), is even less well understood. The concept of this hypothetical form of energy  which exerts negative pressure, appears to explain the dynamics of the present accelerating Universe \cite{Peebles_2003,COPELAND_2006}. According to the thermodynamics, dissipative processes are a universal property of any realistic fluid dynamics. Several 
	cosmological analyses have been carried out considering  dissipative processes \cite{Kamenshchik:2001cp,Fabris2002,PhysRevD.66.043507,Chakraborty:2019con}. 
	Such processes provide both shear  and bulk viscosity in the stress-energy tensor of the cosmic fluid \cite{wang36,1979AnPhy.118..341I,Brevik:2005bj,DiPrisco:2000dw,Avelino:2008ph,Szydlowski:2006ma,Cataldo:2005qh}. However, the shear viscosity can be ignored at large scales for an isotropic and homogeneous Universe. The effect of bulk viscosity  has been widely studied in   viscous dark energy \cite{wang2017,Meng:2008dt,Hu:2005fu,Ren:2005nw,Ren:2006en,Meng:2005jy, 1966ApJ...145..544H,PADMANABHAN1987433,doi:10.1142/S0218271817300245,PhysRevD.95.103509,Anand_2017,Natwariya2020}, viscous dark matter \cite{S0217732317500262,PhysRevD.86.083501,Zimdahl:2000zm,Hipolito-Ricaldi:2009xbk,Ricaldi:2010wrq}, and cosmic inflation  \cite{Bamba:2015sxa}. 
	Viscosity has further been proposed to account for the current acceleration in various models \cite{Gagnon_2011,Das2012,PhysRevLett.114.091301,Mohan2017}. In the presence of viscosity, not only does the Hubble evolution get modified \cite{wang2017}, but  additional entropy is produced which heats up both the baryon and dark matter fluid and alters the thermal evolution of the Universe as well \cite{PhysRevD.100.063539}.

	In this paper, we investigate the effect of the viscous dark energy on the brightness temperature of global 21-cm signal. Our motivation is to study  the comparative impact of the viscous dark energy fluid  vis-a-vis other proposed phenomena such as dark matter decay and annihilation, baryon-dark matter scattering and evaporating primordial black holes that may help us in understanding the possible lowering of the 21-cm temperature \cite{edges}. In the present analysis, we adopt three viscous dark energy (VDE) models in our analysis to explore their effects on the thermal evolution of the Universe in the context of the global signal of 21-cm absorption spectra, and further estimate the bound on such VDE model parameters. The Model I deals with the viscous flow of dark energy only, while in the case of VDE Model II and III, the variation of the parameters $\omega_{\rm de}$ ({\it i.e.}, equation of state parameter of dark energy) and $\Omega_{k,0}$ (current abundance of curvature) are also considered respectively, besides the bulk viscosity of dark energy. 
	We perform an analysis of the interplay of the effect of VDE,
	primordial black hole (PBH) evaporation, particle dark matter decay and annihilation,  and the baryon-dark matter interaction on the 21-cm temperature. We consider PBHs with near present era evaporating mass range $\sim 10^{14} \leq \mathcal{M}_{\rm BH}\lessapprox 2\times 10^{15}$ g, along with a widely used form of the baryon - dark matter scattering cross-section \cite{munoz,upala,21cm_jan,21cm_feb,21cm_mar,21cm_upala,Mahdawi_2018,rennan_3GeV}.

	The paper is organized as follows. Section~\ref{sec:21cm} contains a
	brief overview of the relation of the brightness temperature of the 21-cm absorption line with the thermal evolution of the Universe. In section~\ref{sec:vis_DE}, we describe three viscous dark energy models and their role in gas heating and cosmic evolution. Section~\ref{sec:PBH} describes briefly  the effect of PBHs evaporation in the form of Hawking radiation. Section~\ref{sec:DM_ann_dec} discusses  baryon heating due to particle dark matter annihilation and decay. The effect of baryon-dark matter scattering is discussed in Section~\ref{sec:baryon_DM}. In Section~\ref{sec:T_evol}, the formalism of the thermal evolution is described, where the viscous flow of dark energy, along with the effect of baryon-DM interaction and PBH evaporation are considered.  The results of our analysis are presented in Section~\ref{sec:result}. Finally in Section~\ref{sec:conc}, 
	we present a summary of our main results along with some concluding remarks.

	\section{\label{sec:21cm} 21-cm Brightness Temperature}
	As already mentioned in \autoref{sec:intro}, the 21-cm hydrogen absorption line originated as an outcome of electron transition between the singlet and triplet states of hydrogen atoms. The 21-cm brightness temperature measures the intensity of the said 21-cm spectra, which depends on the optical depth of the medium $\tau$, the spin temperature $T_s$ and the radiation temperature $T_{\gamma}$. The expression of the 21-cm brightness temperature at redshift ($z$) is given by,
	\begin{equation}
		T_{21}=\dfrac{T_s-T_{\gamma}}{1+z}\left(1-e^{-\tau}\right),
		\label{eq:t21}
	\end{equation}
	where, $\tau$ essentially depends on Hubble parameter $H(z)$ as
	\begin{equation}
		\tau = \dfrac{3}{32 \pi}\dfrac{T_{\star}}{T_s}n_{\rm HI} \lambda_{21}^3\dfrac{A_{10}}{H(z)+(1+z)\delta_r v_r}
		\label{eq:tau}
	\end{equation}
	where, $\lambda_{21}\approx21$ cm, 21-cm temperature $T_{\star}=hc/k_B \lambda_{21}=0.068$ K, $A_{10}=2.85\times 10^{-15}\,
	{\rm s^{-1}}$) is the Einstein coefficient \cite{yacine} and $\delta_r v_r$ is the gradient of peculiar velocity.
	
	The spin temperature $T_s$ measures the number ratio of the excited state ($n_1$) to ground state ($n_0$) hydrogen atom as $n_1/n_0=3\exp(-T_{\ast}/T_s)$. In equilibrium the expression of spin temperature $T_s$ can be written as
	\begin{equation}
		T_s = \dfrac{T_{\gamma}+y_c T_b+y_{\rm Ly\alpha} T_{\rm Ly\alpha}}
		{1+y_c+y_{\rm Ly\alpha}},
		\label{eq:tspin}
	\end{equation}
	where,  $T_{\rm Ly\alpha}$ is the Lyman-$\alpha$ temperature, $T_b$ is the baryon temperature,    and $y_{\rm Ly\alpha}$ the  Wouthuysen-Field effect 
	\cite{salpha,jalpha,BH_21cm_2,Yuan_2010,Kuhlen_2006,Yang_2019,21cm_jan}  while $y_c$ is the collisional coupling \cite{BH_21cm_1}, given by
	\begin{equation}
		y_c=\frac{C_{10}T_{\star}}{A_{10} T_b},
	\end{equation}
	with $C_{10}$ being the collision deexcitation rate of the hyperfine level.
	
	Lyman-$\alpha$ photons have a significant contribution to the spin temperature and hence the brightness temperature during the cosmic dawn ($z \lesssim 25$). After recombination, the background photons contribute to flip the spin state of the neutral hydrogen atoms. As a result, the spin temperature ($T_s$) became closer to the $T_{\gamma}$. But later ($z \lesssim 25$), the Lyman-$\alpha$ photons from the newborn stars lead to a quick transition of the spin temperature $T_s=T_b$. Therefore, the spin temperature $T_s$ almost matches 
	with the baryon temperature for $z \lesssim 20$ \cite{tstb2,tstb1} (\autoref{fig:tspin_t21}(a)). This cosmic phenomenon is known as the Wouthuysen-Field effect, which essentially depends on the scattering rate of the Lyman-$\alpha$ photons in the IGM \cite{lya001}.

	\section{\label{sec:vis_DE} Viscous Dark Energy Models}  
	The presence of viscosity in the dark energy fluid may exhibit remarkable effects in the dynamics of the Universe \cite{Wang_2016,1966ApJ...145..544H,PADMANABHAN1987433,doi:10.1142/S0218271817300245,PhysRevD.95.103509,Anand_2017,Natwariya2020}, 
	and in particular, the global 21-cm scenario. In the present analysis we 
	choose an ansatz for the  bulk viscosity of the dark energy to scale as $\zeta(z)=\eta H(z)$  following the work of Wang \emph{et~al.} \cite{wang2017}, where $H(z)$ is the Hubble parameter at redshift $z$, and $\eta$ is a dimensionless proportionality constant in geometrized units (where the velocity of light in 
	vacuum $c=1$ and gravitational constant $G=1$). The effective pressure of the viscous dark energy fluid can be expressed as \cite{wang2017,wang36}
	\begin{equation}
		\bar{p_{\rm de}}=\omega_{\rm de} \rho_{\rm de}-3\zeta H,
		\label{eq:eff_pre}
	\end{equation}
	where $\omega_{\rm de}$ and $\rho_{\rm de}$ are the equation of state parameter and density of dark energy, respectively. 
	Here we investigate the effects of three  Viscous Dark Energy (VDE) models \cite{wang2017}. 
	\begin{itemize}
		\item {\bf Model I (V$\Lambda$DE)} is a one-parameter extended standard $\Lambda$CDM model. In this model, $\omega_{\rm de}$ is fixed at $-1$, so the effective pressure can be written as $\bar{p_{\rm de}}=-\rho_{\rm de}-3\eta H^2$.
		\item {\bf Model II (V$\omega$DE)} has another free parameter, {\it i.e.}, $\omega_{\rm de}$ in addition to the viscosity parameter $\eta$.
		\item {\bf Model III (VKDE)} is similar to the Model I ($\omega_{\rm de}=-1$), but here the additional contribution of curvature is also incorporated. In this particular model the present value of the curvature density parameter $\Omega_k$ is considered as another model parameter.  
	\end{itemize}
	
\begin{figure*}
		\centering
		\begin{tabular}{cc}
			\includegraphics[trim=0 30 30 40, clip, width=0.48\textwidth]{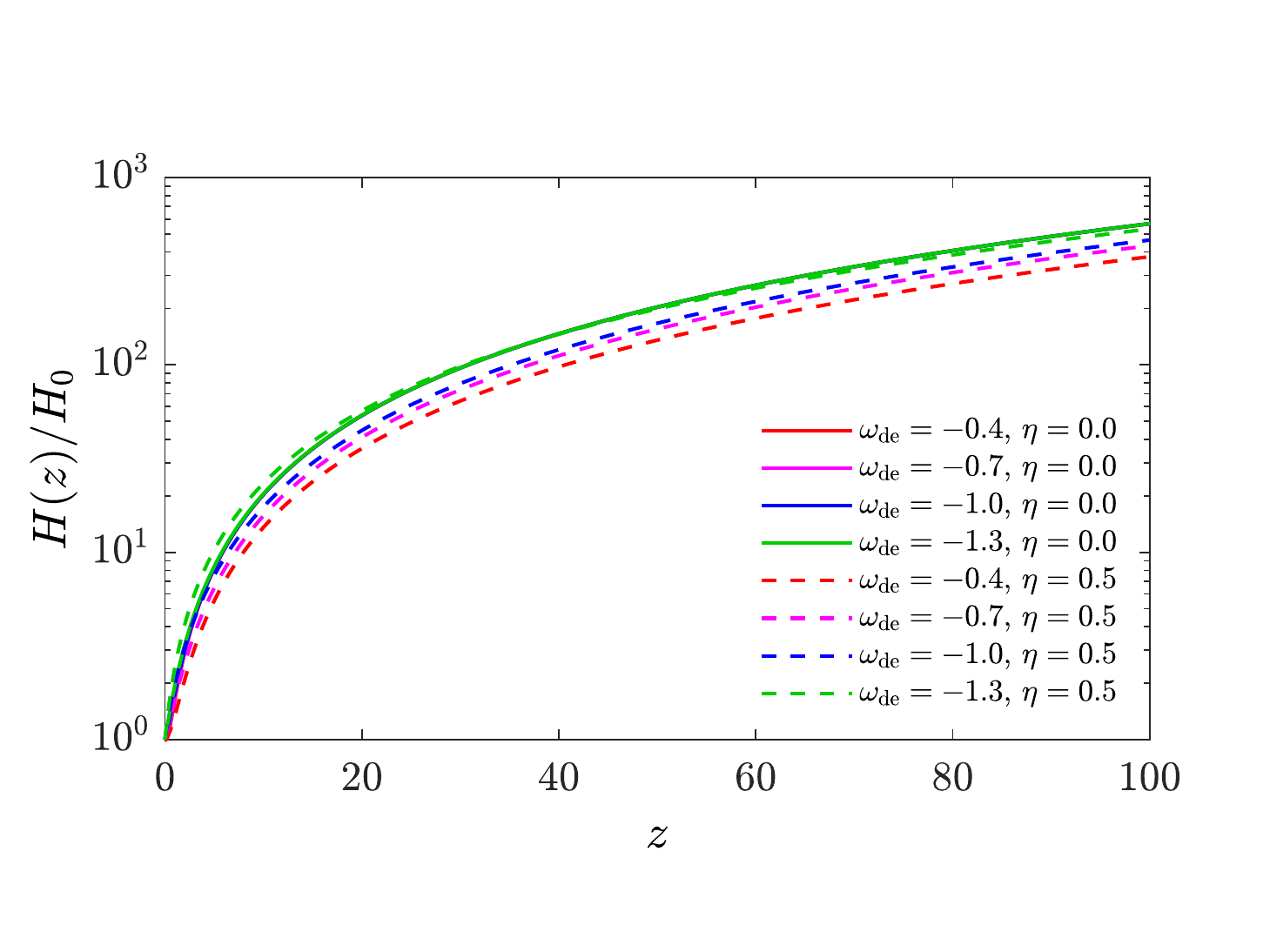}&
			\includegraphics[trim=0 30 30 40, clip, width=0.48\textwidth]{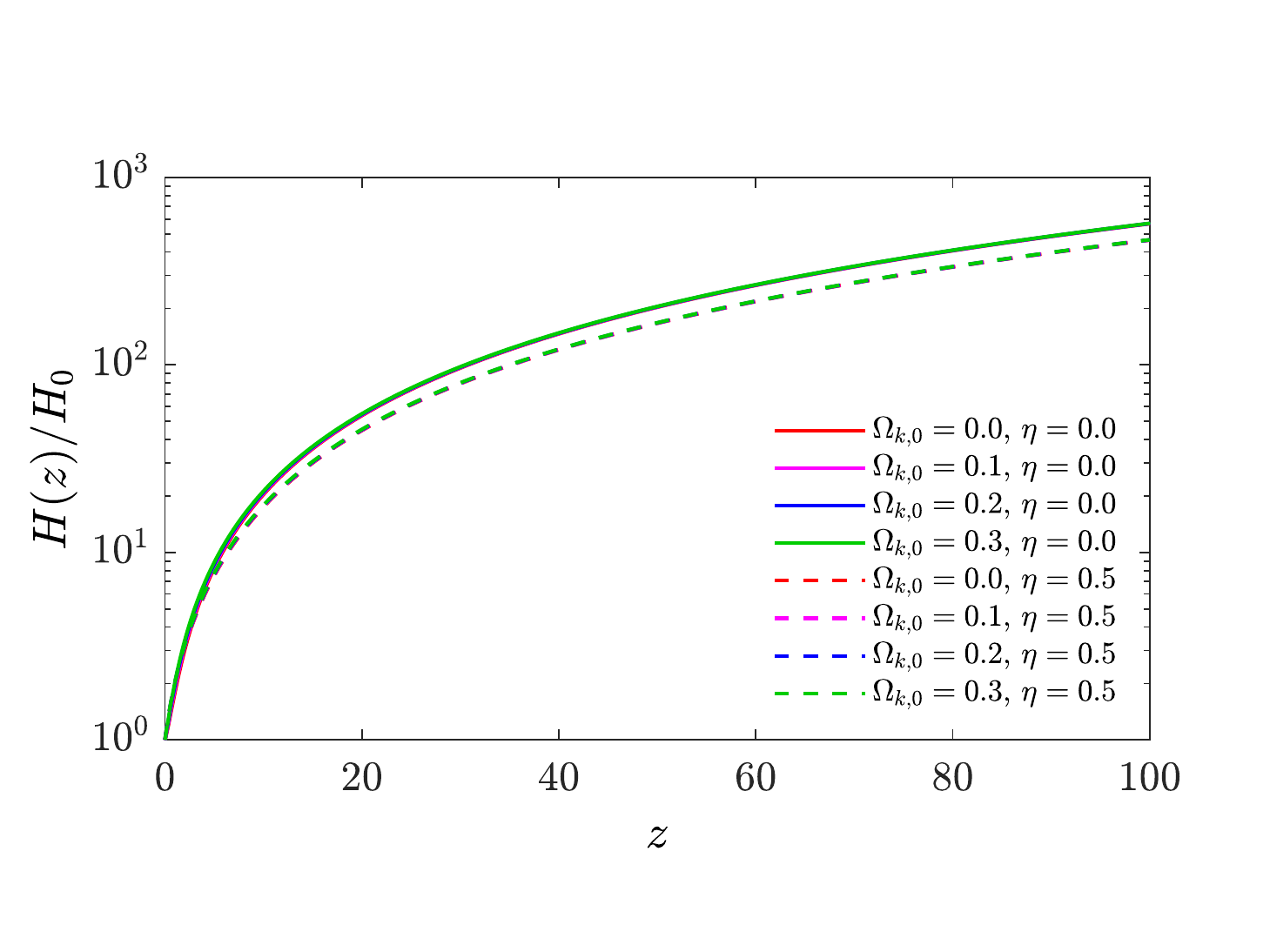}\\
			(a)&(b)\\
			\includegraphics[trim=0 30 30 40, clip, width=0.48\textwidth]{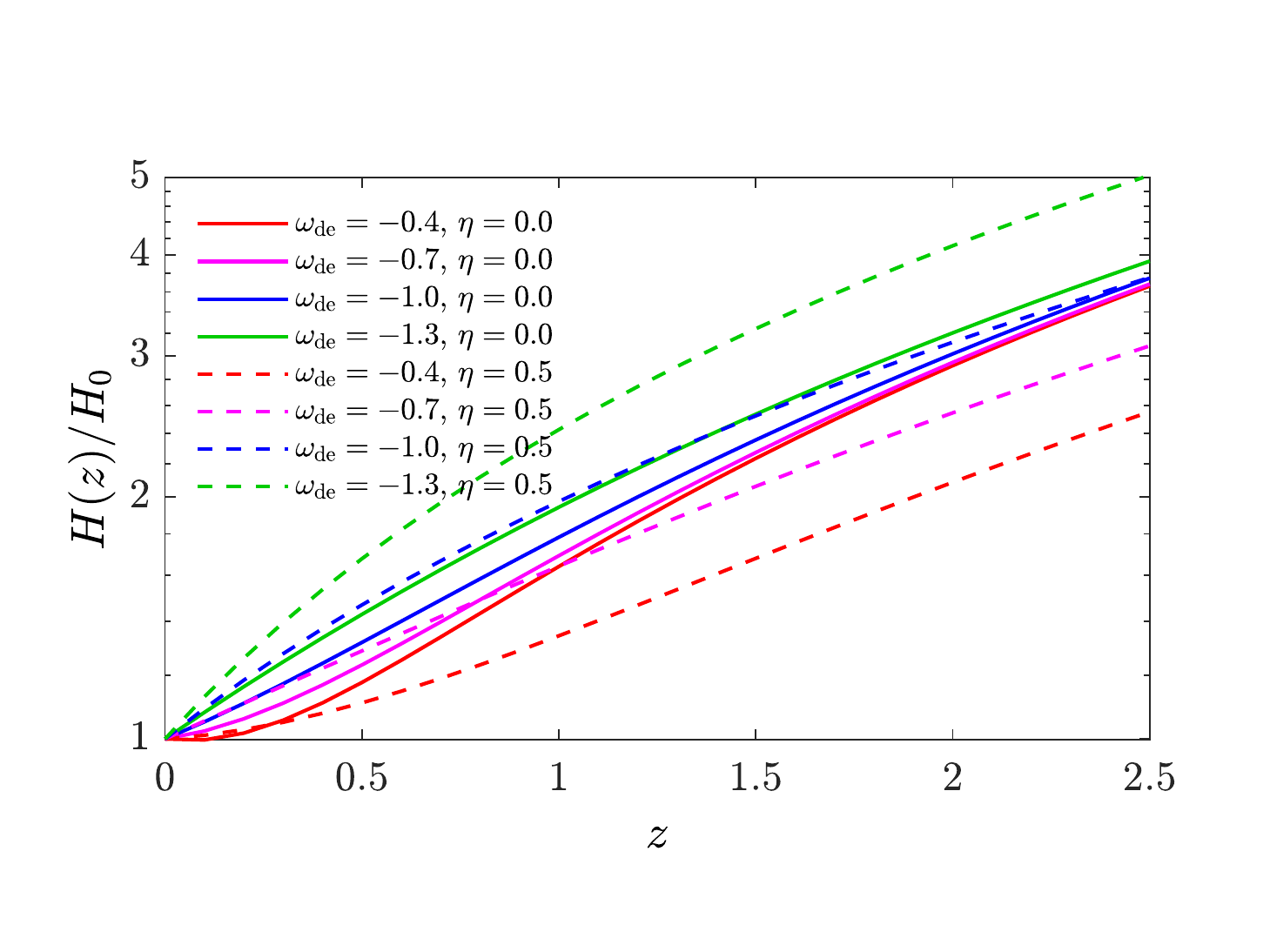}&
			\includegraphics[trim=0 30 30 40, clip, width=0.48\textwidth]{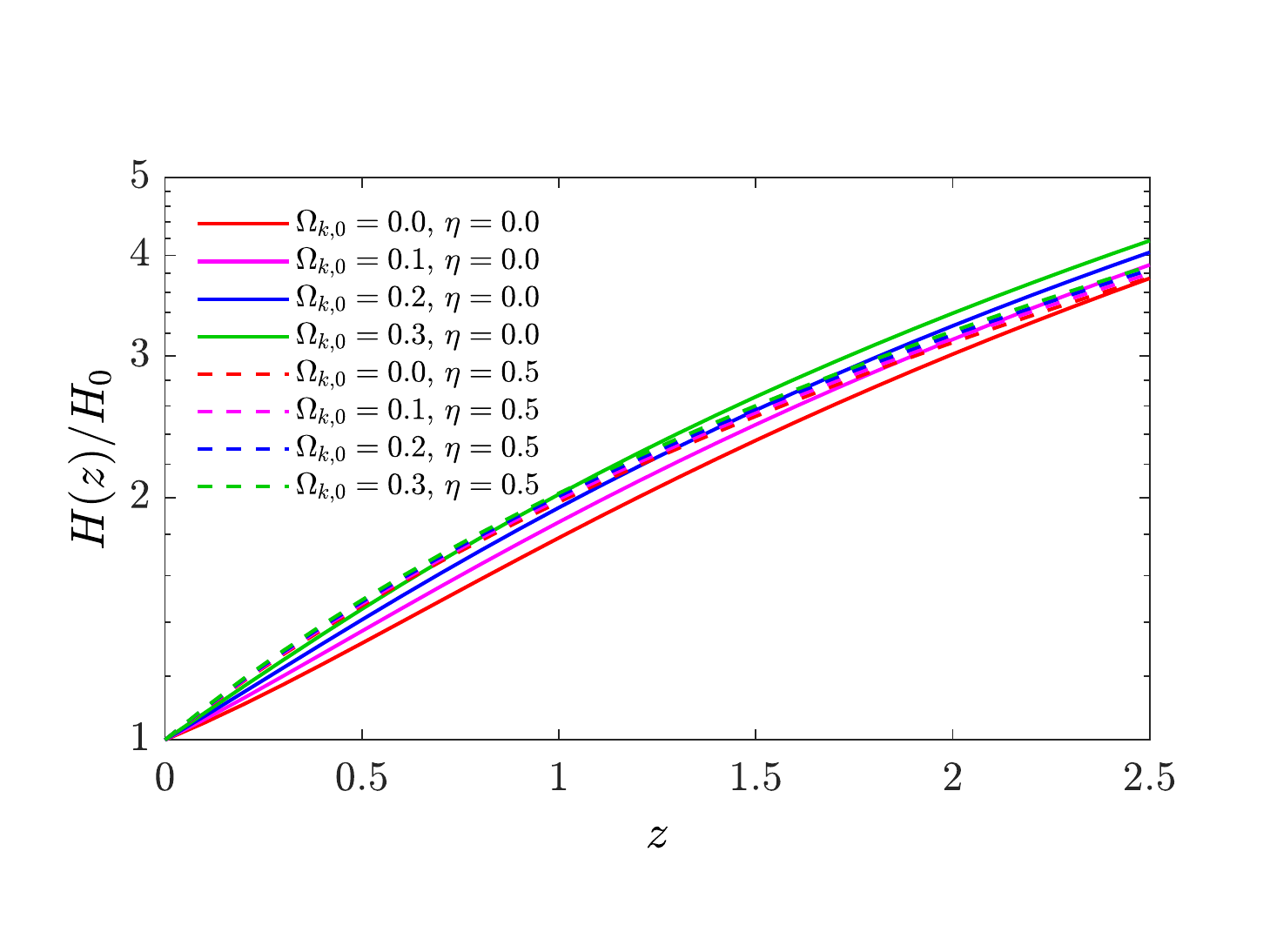}\\
			(c)&(d)\\
			\includegraphics[trim=0 30 10 60, clip, width=0.48\textwidth]{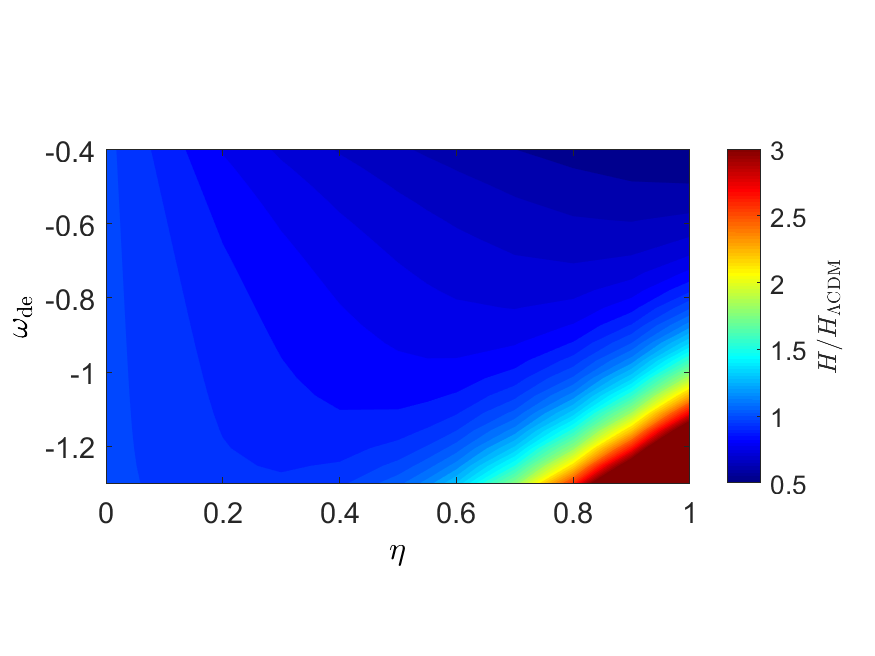}&
			\includegraphics[trim=0 30 10 60, clip, width=0.48\textwidth]{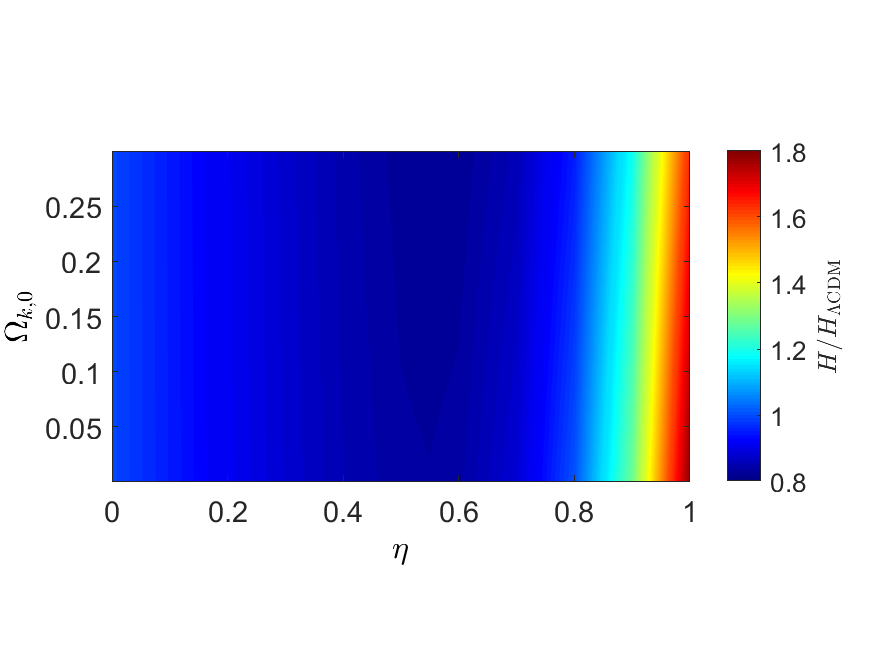}\\
			(e)&(f)\\
		\end{tabular}
		\caption{\label{fig:Hmod2_3} \autoref{fig:Hmod2_3}(a) and (b) show the variations of Hubble parameter with different VDE model parameters for Model II and Model III respectively. \autoref{fig:Hmod2_3}(c) and (d) are the zoomed views of \autoref{fig:Hmod2_3}(a) and (b) respectively, at lower redshifts. Contour representations of Hubble parameter for VDE Model II and Model III at $z=17.2$ are furnished in \autoref{fig:Hmod2_3}(e) and (f) respectively.}
	\end{figure*}

	At low redshifted epochs, the effect of bulk viscosity is small, and as a result, the evolution of the Universe is very close to the standard cosmology ($\Lambda$CDM model) for lower redshifts. All three VDE models discussed above deal with bulk viscosity of dark energy. However, in the second model an additional effect of dark energy equation of state parameter $\omega_{\rm de}$ is studied, which provides a comparatively faster expansion of the Universe \cite{wang2017} at low redshifts  for $\omega_{\rm de} \leq -1$. However, for $\omega_{\rm de} \gtrapprox -0.7$ the expansion rate decreases significantly (see also  \autoref{fig:Hmod2_3}). On the other hand, the effect of spatial curvature of the Universe is investigated in presence of viscous flow of DE in the VDE Model III. Applying the latest bounds of curvature \cite{planck}, it can be seen that the evolution of the Universe is minimally modified at  late times  \cite{wang2017}.	
	
	The Hubble parameter for all three VDE models can be obtained from 
	the Friedmann equations and \autoref{eq:eff_pre}, given by
	\begin{eqnarray}
		H(z) &=& H_0 \left[\dfrac{1}{1+\eta}\Omega_{m,0} (1+z)^3 + 
		\left(1-\dfrac{1}{1+\eta}\Omega_{m,0} \right) (1+z)^{-3\eta}\right]^{1/2},\label{eq:mod1}\\
		H(z) &=& H_0 \left[\dfrac{\omega_{\rm de}}{\omega_{\rm de}-\eta}\Omega_{m,0} (1+z)^3 + \left(1-\dfrac{\omega_{\rm de}}{\omega_{\rm de}-\eta}\Omega_{m,0} \right) (1+z)^{3(1+\omega_{\rm de}-\eta)}\right]^{1/2},\label{eq:mod2}\\
		H(z) &=& H_0 \left[\dfrac{2}{2+3\eta}\Omega_{k,0} (1+z)^2 + 
		\dfrac{1}{1+\eta}\Omega_{m,0} (1+z)^3 + \right.\nonumber\\ &&\left.\left(1-\dfrac{2}{2+3\eta}\Omega_{k,0}-\dfrac{1}{1+\eta}\Omega_{m,0} \right) (1+z)^{-3\eta}\right]^{1/2}\label{eq:mod3}.
	\end{eqnarray}
	In the above expressions, $H_0$ denotes the present value of the Hubble parameter, $\Omega_{m,0}$ and $\Omega_{k,0}$ are the current density parameters of matter (baryonic matter + dark matter) and curvature, respectively. The constraints and parameters for individual models are tabulated in Table~\ref{tab_mod}.
	\begin{table}
		\centering
		\begin{tabular}{ccc}
			\hline
			Model & Constraints & Free parameters\\
			\hline
			I & $\quad$ $\omega_{\rm de}=-1$, $\Omega_k=0$ $\quad$ & $\eta$\\
			II & $\Omega_k=0$ & $\eta$, $\omega_{\rm de}$\\
			III & $\omega_{\rm de}=-1$ & $\eta$, $\Omega_k$\\
			\hline
		\end{tabular}
		\caption{\label{tab_mod} Constraints and parameters of three viscous dark energy models}
	\end{table}
	The generalized expression of the Hubble parameter for all the three models (\autoref{eq:mod1}, \autoref{eq:mod2}, \autoref{eq:mod3}) can be written as,
	\begin{eqnarray}
		H(z)&=&H_0 \left[\dfrac{2}{2+3\eta}\Omega_{k,0} (1+z)^2 +
		\dfrac{\omega_{\rm de}}{\omega_{\rm de}-\eta}\Omega_{m,0} (1+z)^3\right.\nonumber \\
		&&\left. + \left(1-\dfrac{2}{2+3\eta}\Omega_{k,0}-\dfrac{\omega_{\rm de}}{\omega_{\rm de}-\eta}\Omega_{m,0} \right) (1+z)^{3(1+\omega_{\rm de}-\eta)}\right]^{1/2}.
		\label{eq:general_H}
	\end{eqnarray}

	\begin{figure*}
		\centering
		\begin{tabular}{cc}
			\includegraphics[width=0.48\textwidth]{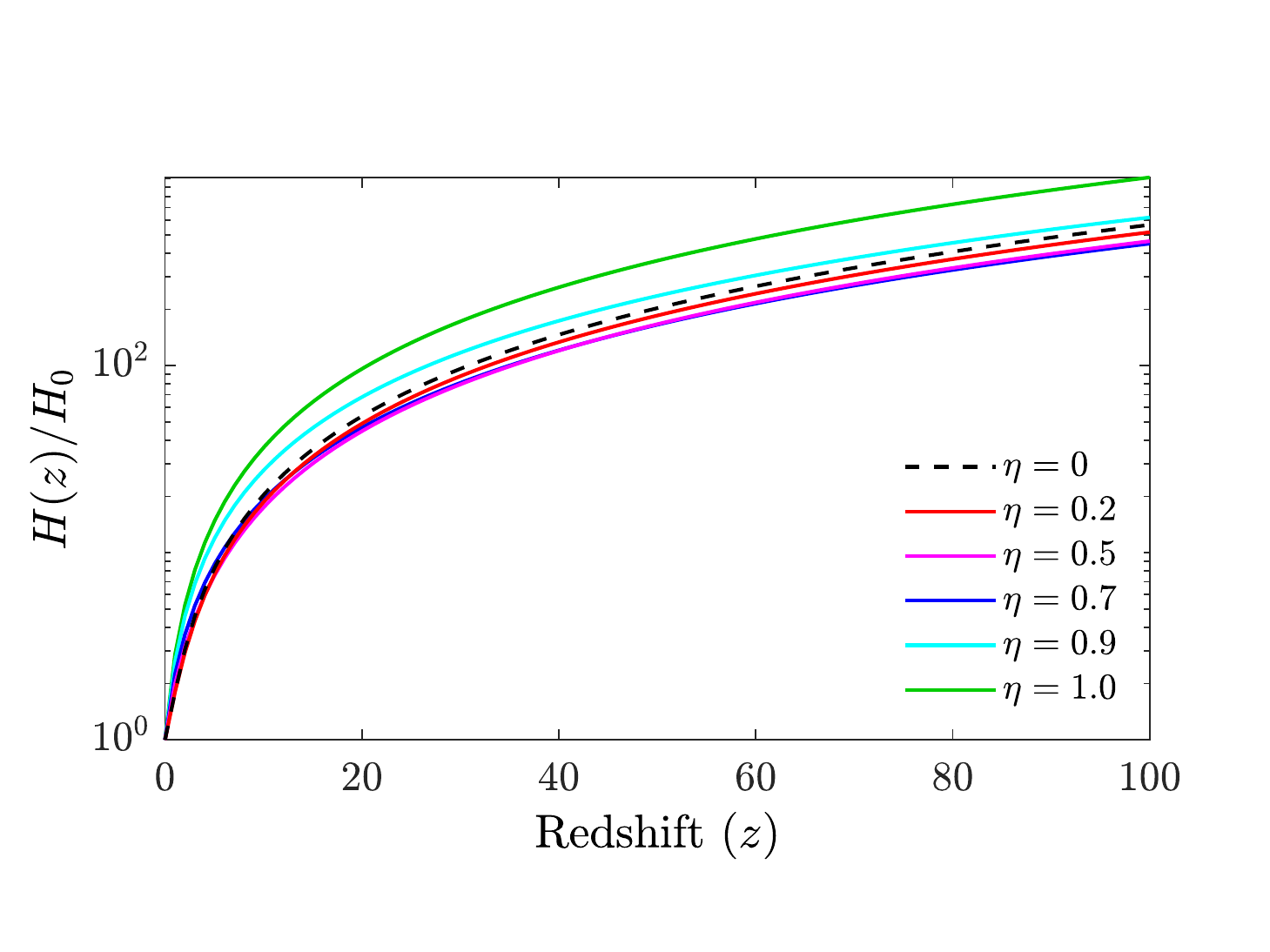}&
			\includegraphics[width=0.48\textwidth]{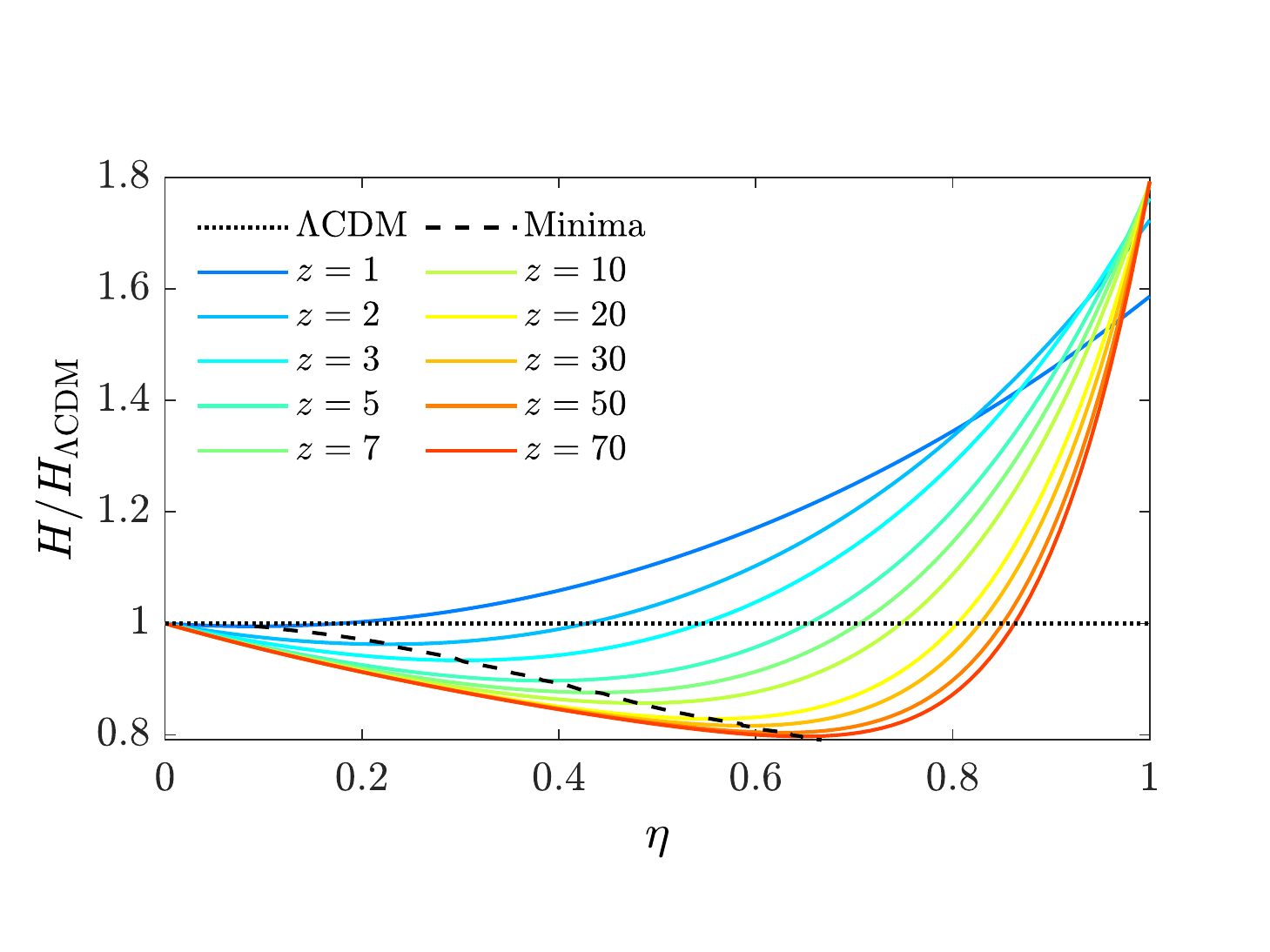}\\
			(a)&(b)\\
		\end{tabular}
		\caption{\label{fig:Hmod1}(a) The variation of the dimensionless Hubble parameter ($H(z)/H_0$) with redshift $z$ for different values of $\eta$ for the case of VDE Model I. (b) Variation of the Hubble parameter with $\eta$ (VDE Model I) at different fixed values of $z$, where the black dashed line denotes the Hubble parameter for the $\Lambda$CDM model.  For every chosen value of $z$, there is a minimum value
			of $H/H_{\rm \Lambda CDM}$ with respect to $\eta$.
			The positions of these minimum points for different $z$ are shown by the black dashed line.}
	\end{figure*}
	
	In presence of the viscous flow of dark energy, the effective pressure of DE modifies remarkably (see \autoref{eq:eff_pre}). As a result, a significant departure may be observed in the evolution of the Hubble parameter
	compared to the $\Lambda$CDM model. The variation of the dimensionless Hubble parameter in presence of viscosity with different values of $\eta$ is graphically represented in \autoref{fig:Hmod1}(a). From this figure, it can be seen that the variation of $H(z)/H_0$ is not linear with $\eta$.
	
	The variation of $H(z)$ with $\eta$ for any particular redshift $z$ is described in \autoref{fig:Hmod1}(b), where the Hubble parameter is written in the form of $H/H_{\rm \Lambda CDM}$, $H_{\rm \Lambda CDM}$ being the value of Hubble parameter for the $\Lambda$CDM model (represented by the black dashed line). From \autoref{fig:Hmod1}(b) it can be noticed that for a fixed redshift
	$z$, the Hubble parameter at first decreases with increasing $\eta$. 
	However, there is a minimum value of $H$ for every value of $z$ beyond 
	which $H$ increases rapidly with $\eta$. These minimum points for different redshift $z$ are depicted through the black dashed curve. The minima points are located at higher values of $\eta$ for higher values of $z$, but tend toward $\eta=0$ and $H/H_{\rm \Lambda CDM}=1$ at the lower redshifted epochs.
	
	Similar variations in the case of VDE Model II and III are displayed in \autoref{fig:Hmod2_3}(a) and (b), where the model parameters are $\omega_{\rm de}$, $\eta$ and $\Omega_{k,0}$, $\eta$ respectively. \autoref{fig:Hmod2_3}(c) and (d) are zoomed view of \autoref{fig:Hmod2_3}(a) and (b), respectively, which reveal the behavior of $H$ in lower redshifted epochs for Model II and Model III. \autoref{fig:Hmod2_3}(a) provides a consistency check for our calculations, as it can be clearly observed that for $\eta=0$ (DE model without viscosity) the variation of $H(z)$ with $\omega_{\rm de}$ is negligible
	at large redshifts, although at lower values of $z$ significant variation is obtained (see \autoref{fig:Hmod2_3}(c)). On the other hand, for VDE models,
	{\it i.e.,} $\eta=0.5$ (or any non-zero value of $\eta$) the variation of $H$ with $\omega_{\rm de}$ is remarkable. The variation with $\Omega_{k,0}$ in the case of Model III is shown in \autoref{fig:Hmod2_3}(b) and \autoref{fig:Hmod2_3}(d). From these figure (\autoref{fig:Hmod2_3}(b) and (d)) one can see that the variation with $\Omega_{k,0}$ is negligible, but a small variation can be observed for $z<2.5$ (see \autoref{fig:Hmod2_3}(d)).

	We further explore the variation of the Hubble parameter with 
	respect to the model parameters at a fixed value of $z$. Here we display
	our results  for $z=17.2$ which  corresponds to the EDGES observation. In \autoref{fig:Hmod2_3}(e) the contour representation is plotted for the Hubble parameter in $\omega_{\rm de}$-$\eta$ parameter space in the case of the VDE Model II at $z=17.2$. From this plot it can be observed that for $-0.9 \lessapprox \omega_{\rm de} \lessapprox -0.4$, the value of $H/H_{\rm \Lambda CDM}$ (the departure of the Hubble parameter from the $\Lambda$CDM value) is
	rather small throughout the range of viscosity $\eta$. However, in the case of phantom dark energy ({\it i.e.} $\omega_{\rm de} < -1$) \cite{Bouali_2019,Ludwick_2017}, the  values of $H/H_{\rm \Lambda CDM}$ can be significantly large for $\eta\gtrapprox 0.6$. As a consequence, the late behavior of the Hubble evolution can be  sharply distinguished for such cases. 
	On the other hand, for the VDE Model III, the variation is essentially governed by $\eta$ only, which is graphically represented in \autoref{fig:Hmod2_3}(f). Although, here a minute variation of the Hubble parameter is observed with respect to the curvature $\Omega_{k,0}$, the dependence on $\eta$ may be prominent for higher values of viscosity.

	The viscosity of the dark energy fluid does not only affect the Hubble evolution of the Universe, but an additional heat is also produced as an outcome of the viscous flow. The entropy generated due to viscous flow is discussed in Ref.~\cite{weinberg1972gravitation,PhysRevD.100.063539}. 
	Applying the formalism of Ref.~\cite{weinberg1972gravitation,PhysRevD.100.063539}, the amount of entropy ($S$) produced per unit volume in the FLRW metric is given by 
	\begin{equation}
		\nabla_{\mu}S^{\mu}=\dfrac{\zeta}{T_{\rm de}} \left(\nabla_{\mu}u^{\mu}\right)^2,
	\end{equation}
	where $T_{\rm de}$ is the effective temperature of the viscous dark energy fluid, and $u^{\mu}$ represents the four velocity of the viscous fluid.	Now, using the second law of thermodynamics, the amount of heat produced $Q$ per unit volume $V$, per unit time in the comoving frame due to viscous flow of the dark energy fluid can be expressed as,
	\begin{equation}
		\dfrac{{\rm d}Q}{{\rm d}V{\rm d}t}=\zeta \left(3H(z)\right)^2.
		\label{eq:dqdvdt}
	\end{equation}
	This amount of entropy heats up the matter (both baryons and dark matter), 
	which could lead  to an increase the spin temperature. On the other hand, it also helps to cool down the baryons faster than the usual, as an outcome of the modified Hubble evolution.
	In our subsequent analysis we perform a study of the interplay of these
	competing effects on the 21-cm brightness temperature.
	
	\section{\label{sec:PBH} Effect of Primordial Black Holes}  
	Besides the viscous flow of dark energy, the energy injection in the form of Hawking radiation from PBHs could modify remarkably the global 21-cm signature \cite{BH_21cm_0,BH_21cm_1,21cm_feb}. Primordial black holes \cite{khlopov_1,khlopov_2,khlopov_3,juan} are believed to be formed due to the collapse of overdensity zones in the early ages of the Universe. The overdensity zones are characterized by the size, which should be greater than the Jeans length $R_j = \sqrt{\displaystyle\frac {1} {3G \rho}}$ \cite{roos2004introduction}.  PBHs are produced when $\delta_{\rm min} \leq \delta$, where  $\delta$ denotes the density contrast and $\delta_{\rm min}$ is the lower bound of the density contrast with the density $\rho = \rho_c + \delta \rho$, $\rho_c$ being the critical density for collapse and $\delta_{\rm min}$  the threshold of PBH formation. Several mechanisms have been proposed to explain the formation of PBHs \cite{stpbh_1,stpbh_2,stpbh_3,stpbh_4,stpbh_5,stpbh_6,stpbh_7,fsc_1,fsc_2,fsc_3,ccs_1,ccs_2,ccs_3}. 
	
	Primordial black holes come with a wide range of mass \cite{Zhou:2021tvp,BH_21cm_2,BH_21cm_3,BH_21cm_1}. Primordial black holes having evaporation time scale longer than age of the Universe at the time of recombination are relevant to study the post-recombination energy injection. The corresponding mass range for such PBHs is $\sim 10^{14} \leq \mathcal{M}_{\rm BH}\lessapprox 2\times 10^{15}$ g. In addition, PBHs with mass $\gtrapprox 10^{15}$ g can survive to today, and such PBHs are subject to study CMB damping constraints \cite{cmbdamp,cmbdamp1,cmbdamp2} and indirect searches for extragalactic cosmic rays \cite{carr}. Consequently, in several recent works, PBHs having the above mass range have been considered in order to estimate the constraints on the abundance of PBHs \cite{21cm_feb,BH_21cm_2} in the context of 21-cm scenario, as PBHs with this mass range are likely to impact the 21-cm signal most. In the present analysis we consider PBHs of masses $\sim 10^{14} \leq \mathcal{M}_{\rm BH}\lessapprox 2\times 10^{15}$ g.
	
	In the case of a PBH having mass $M_{\rm{BH}}$, the rate of mass evaporation in the form of Hawking radiation can be approximated as \cite{BH_F}
	\begin{equation}
		\dfrac{{\rm d}M_{\rm{BH}}}{{\rm d}t} \approx -5.34\times10^{25} \left(\sum_{i} \mathcal{F}_i\right) \left(\dfrac{M_{\rm{BH}}}{\rm g}\right)^{-2} \,\,\rm{g/sec}
		\label{eq:PBH}
	\end{equation}
	where, $\sum_{i} \mathcal{F}_i$ is the total contribution of different species, which is essentially governed by the black hole temperature $T_{\rm BH}$ ($=1.05753 \times \left(M_{\rm{BH}}/10^{13} {\rm g}\right)^{-1}$ \cite{BH_F}), defined as \cite{BH_F},
	\begin{eqnarray}
		\sum_{i} \mathcal{F}_i&=&1.569+3.414\exp \left(-\frac{0.066}{T_{\rm{BH}}} \right)+
		1.707\exp \left(-\frac{0.413}{T_{\rm{BH}}} \right)+ 
		1.707\exp \left(-\frac{0.11}{T_{\rm{BH}}} \right)\nonumber \\
		&&+1.707\exp \left(-\frac{22}{T_{\rm{BH}}} \right)+
		1.707\exp \left(-\frac{1.17}{T_{\rm{BH}}} \right)+
		0.569 \exp\left(-\frac{0.0234}{T_{\rm{BH}}} \right)\nonumber \\
		&&+0.569\exp \left(-\frac{0.394}{T_{\rm{BH}}} \right)+
		0.963\exp \left(-\frac{0.1}{T_{\rm{BH}}}\right)
	\end{eqnarray}
	In the above equation, the combined contribution of $e^-$, $e^+$, $\nu$ and photons in Hawking radiation is denoted by the first term. The second term represents the contribution of $u$ and $d$ quark, while the third, fourth, fifth and sixth terms are the same for $c$, $s$, $t$ and $b$ quarks respectively. The effect of muon, $\tau$ and gluons are included in the seventh, eighth and ninth terms respectively.
	
	In the present analysis, the chosen mass range of PBHs are significantly lower than stellar mass. As a consequence, alongside the $\gamma$ and electron channels we consider the contributions of other standard model decay channels, which heats up the baryons by further producing photons, electrons and positrons via subsequent cascade decay \cite{PhysRevD.41.3052,chen,carr,BH_21cm_2}.
	
	The total rate of energy injected by PBHs is given by \cite{BH_21cm_2},
	\begin{equation}
		\left.\dfrac{{\rm d} E}{{\rm d}V {\rm d}t}\right|_{\rm{BH}}=-\dfrac{{\rm d} M_{\rm{BH}}}{{\rm d} t} n_{\rm BH}(z),
		\label{eq:dedvdt}
	\end{equation}
	where, the PBH number density $n_{\rm{BH}}(z)$ is given by,
	\begin{eqnarray}
		n_{\rm{BH}}(z)&=&\beta_{\rm BH}\left(\dfrac{1+z}{1+z_{\rm eq}}\right)^3 \dfrac{\rho_{\rm c,eq}}{\mathcal{M}_{\rm BH}} \left(\dfrac{\mathcal{M}_{\rm H,eq}}{\mathcal{M}_{\rm H}}\right)^{1/2} \left(\dfrac{g^i_{\star}}{g^{\rm eq}_{\star}}\right)^{1/12}\nonumber\\
		&\approx&1.46 \times 10^{-4}\beta_{\rm BH} \left(1+z\right)^3 \left(\dfrac{\mathcal{M}_{\rm BH}}{\rm g}\right)^{-3/2} {\rm cm^{-3}}.
		\label{eq:beta}
	\end{eqnarray}
	In the above expression $\beta_{\rm BH}$ and $\mathcal{M}_{\rm BH}$ are the initial mass fraction and initial mass of primordial black holes. The other parameters $\mathcal{M}_{\rm H}$, $g_{\star}^i$ are the horizon mass and the total number of effectively massless degrees of freedom during the formation of PBHs respectively, and $\mathcal{M}_{\rm H,eq}$ and $g_{\star}^{\rm eq}$ are the same at the epoch of matter-radiation equality.
	
	\section{\label{sec:DM_ann_dec} Dark Matter Annihilation and Decay}
	
\begin{figure}
		\centering
		\includegraphics[width=0.6\linewidth]{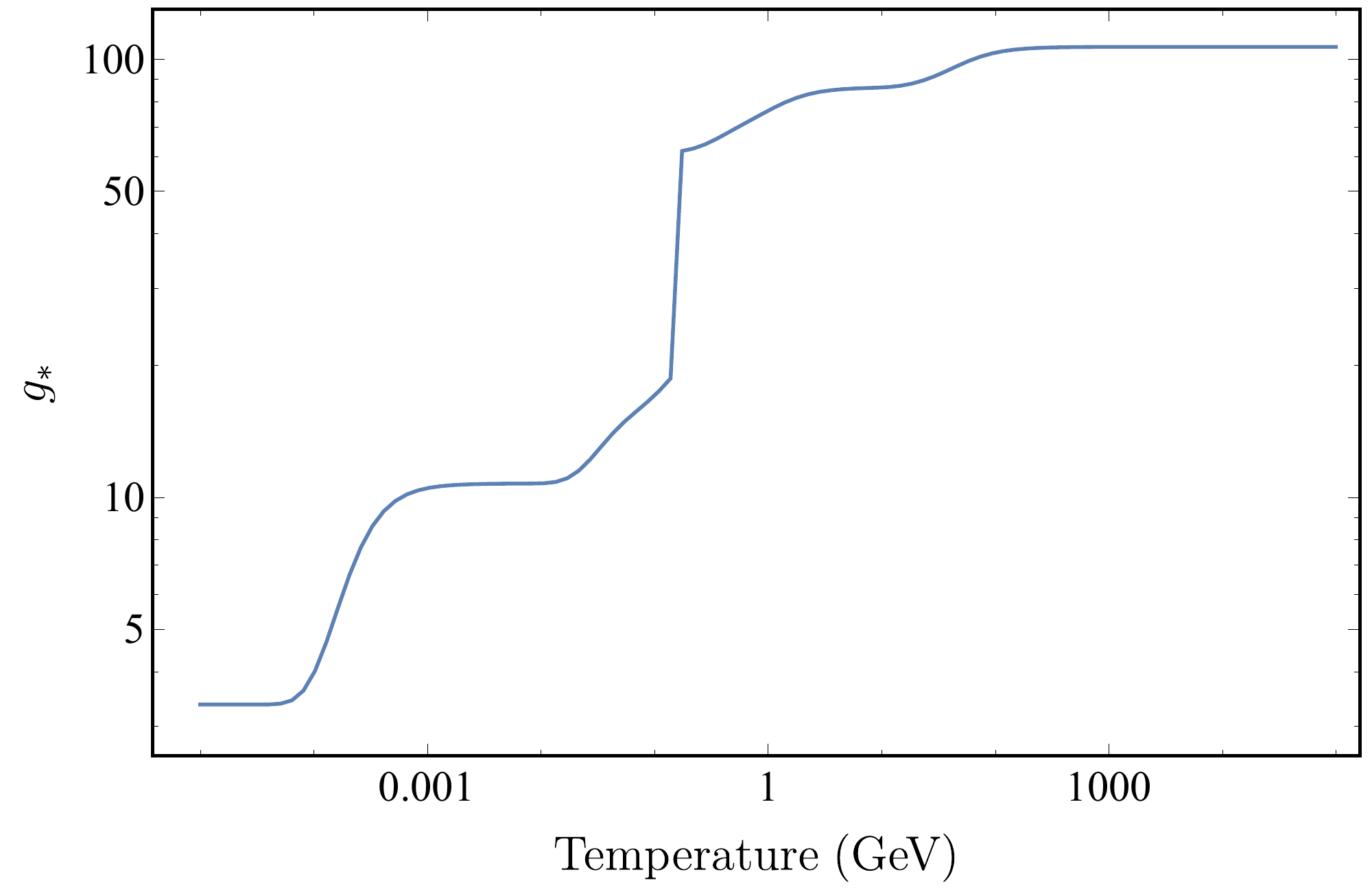}
		\caption{Effective degrees of freedom ($g_*$) in the early Universe.}
		\label{fig:gstr}
	\end{figure}

	Dark matter annihilation and decay are two notable sources of baryon heating. The heating due to these two processes are governed by the velocity averaged annihilation cross-section of DM ($\langle \sigma v \rangle$) and the decay lifetime ($\tau_{\chi}$), respectively. Both these parameters $\langle \sigma v \rangle$ and $\tau_{\chi}$ depend on the DM model and are essentially functions of the dark matter mass $m_{\chi}$. The velocity averaged annihilation cross-section $\langle \sigma v \rangle$ of  dark matter  \cite{IDM_1,IDM_2,Cheng:2002ej,servant_tait,Hooper:2007gi,Majumdar:2003dj,wimpfimp,dsphs}  can be estimated by using the cosmic relic of the dark matter species \cite{GONDOLO1991145,aman}. In order to estimate the cross-section, one needs to first evaluate the relativistic degrees of freedom ($g_*(T)$) as a function of the temperature $T$ (see \autoref{fig:gstr}), given by
	\begin{equation}
		g_*(T) = \sum_{b, f}g_{i}(T),
		\label{eq:gstr}
	\end{equation}
	where $b$ and $f$ denote the contributions of bosons and fermions respectively and $g_i(T)$ is the effective degrees of freedom for the $i^{\rm th}$ particle, given by
	\begin{equation}
		g_i(T) = \dfrac{15 g_i}{\pi^4}x^4\int_1^{\infty} \dfrac{y(y^2-1)^{1/2}}{\exp(x_i y)+\eta_i}y dy.
		\label{eq:gi}
	\end{equation}
	In the above expression $\eta_i$ is a constant which is $1$ for fermions and $-1$ for bosons \cite{GONDOLO1991145}, and $x_i=m_i/T$, where $m_i$ is the mass of the $i^{\rm th}$ species.
	\begin{figure}[b]
		\centering
		\includegraphics[width=0.6\linewidth]{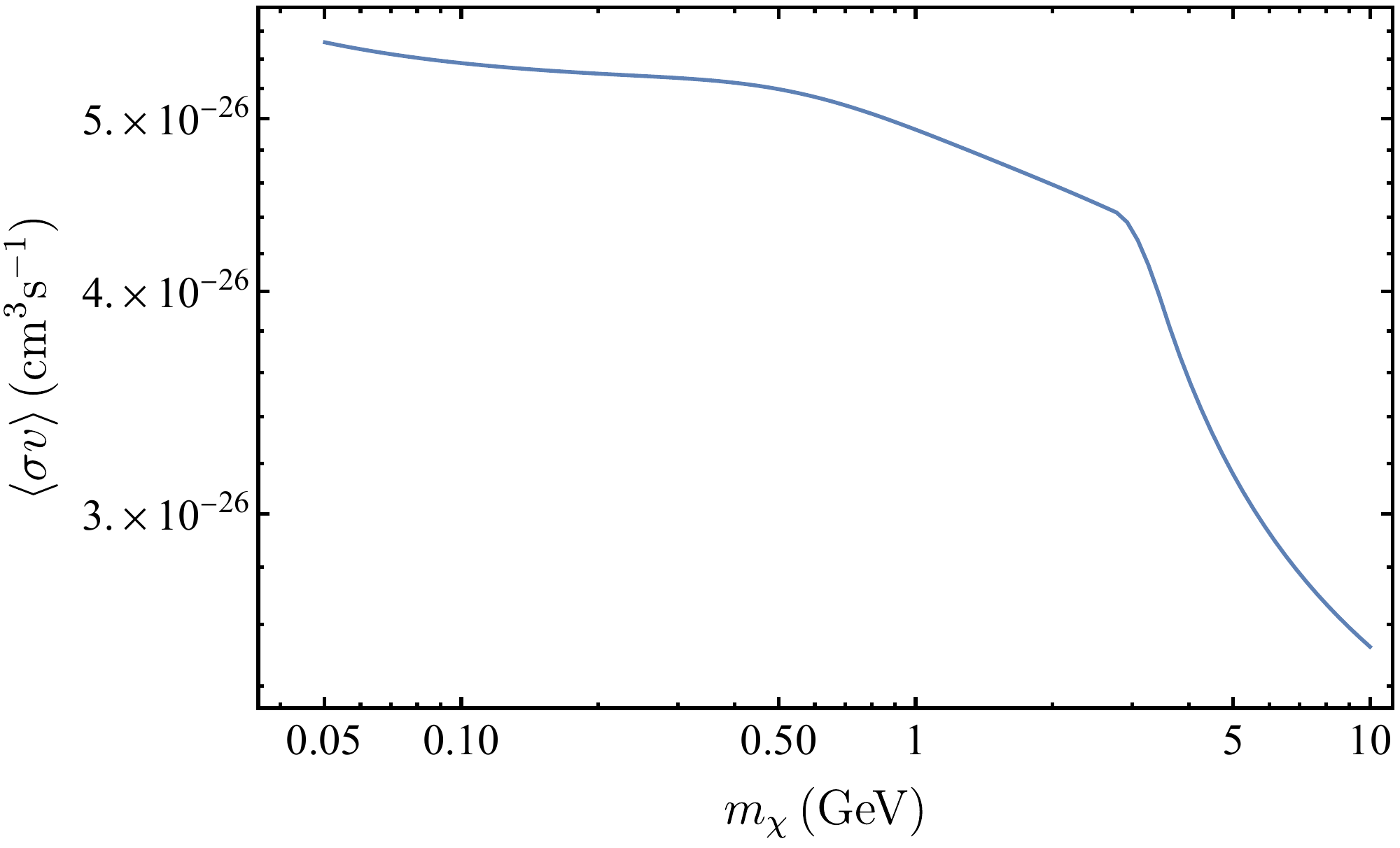}
		\caption{Velocity averaged annihilation cross-section of dark matter.}
		\label{fig:sigmav}
	\end{figure}
	After decoupling of a dark matter candidate, the velocity averaged annihilation cross-section can be estimated by solving  numerically the equation \cite{GONDOLO1991145} (see Fig.~\ref{fig:sigmav})
	\begin{equation}
		\dfrac{1}{Y_0}= \left(\dfrac{45}{\pi}G\right)^{1/2} \int_{T_0}^{T_f}g_*^{1/2}\langle\sigma v\rangle dT,
		\label{eq:gondolo_5.5}
	\end{equation}
	where $T_0$ and $T_f$ are the present photon temperature and the freeze-out temperature of the DM candidate, respectively.
	$Y_0$  denotes the present value of the comoving abundance  and can be written as \cite{GONDOLO1991145}
	\begin{equation}
		\Omega h^2 \theta^{-3}=2.8282\times 10^8 \dfrac{m}{\rm {GeV}} Y_0,
		\label{eq:y0}
	\end{equation}
	where $\theta$ is the CMB temperature in units of 2.75 K, $h$ is the Hubble constant in units of 100 km/s/Mpc and $\Omega$ is the present cosmological density parameter which is adopted from the experimental data of the Planck  experiment \cite{planck}. Now, comparing \autoref{eq:gondolo_5.5} and \autoref{eq:y0} the values of $\langle \sigma v \rangle$ are estimated numerically for different DM masses $m_{\chi}$. In the present analysis we adopt this procedure to estimate $\langle \sigma v \rangle$ for the individual cases. On the other hand, decaying DM particles having lifetime ($\tau_{\chi}$) remarkably longer than the age of the Universe are considered. From several recent analyses it follows that $\tau_{\chi}$ lies within the range of $10^{24}\sim 10^{30}$ sec \cite{BH_21cm_1,083517,063539,dmdec2,Pandey:2019cre}.
	
	The annihilation and decay of dark matter particles produce standard model particles, which heat up and ionize the baryonic matter of the Universe. The rate of energy injected as an outcome of these processes are given by, respectively,
	\begin{eqnarray}
		\left.\dfrac{{\rm d}E}{{\rm d}V{\rm d}t}\right|_{\rm ann.}&=&f_{\chi,{\rm ann}}^2\rho_{\chi}^2 \dfrac{\langle \sigma v \rangle}{m_{\chi}},\\
		\left.\dfrac{{\rm d}E}{{\rm d}V{\rm d}t}\right|_{\rm dec.}&=&f_{\chi,{\rm dec.}}\rho_{\chi} \dfrac{1}{\tau_{\chi}}.
	\end{eqnarray} 
	In the above expressions, $f_{\chi,{\rm ann}}$ and $f_{\chi,{\rm dec}}$ are the mass fractions of dark matter that contribute in annihilation and decay, respectively. In the present analysis we set both quantities to unity,
	following  Ref.~\cite{PhysRevD.98.023501}.

	\section{\label{sec:baryon_DM} Baryon-Dark Matter Scattering}
	
	Apart from the phenomena of dark matter annihilation and decay, we
	consider the effect of baryon-dark matter scattering  in the context of thermal evolution of the Universe. The interaction between baryons and cold dark matter (CDM) plays the key role in the cooling process of baryons, which essentially depends on the baryon-DM relative velocity $V_{\chi b}$ ($=V_{\chi}-V_b$, where $V_b$ and $V_{\chi}$ are the bulk velocities of baryons and dark matter respectively) and baryon-DM cross-section $\sigma$. The relevance of such relative velocity was first pointed out in Ref.~\cite{rel_vel_1st} in the context of small-scale structure formation. After the decoupling of baryons at redshift $z\approx 1010$, baryons start scattering with dark matter particles. Consequently, we start our analysis from $z=1010$ when the relative velocity $V_{\chi b}=V_{\chi b,0}$. The baryon-DM cross-section term is velocity ($v$) dependent and can be expressed using a general expression $\sigma = \left(\sigma_0\right) v^{n}$. The value of $n$ depends on the nature of the dark matter candidates. 
	
	In several recent works, dark matter-baryon interaction cross-section is investigated for a wide range of $n$ \cite{dvorkin2020cosmology,Nadler_2019,Bhoonah_2018,Kovetz_2018,Mack_2007}. $n=\pm 2$ attributes to the case of dark matter candidates having magnetic and/or electric dipole moment. On the other hand, for Yukawa potential \cite{yukawa}, several values of $n$ are applicable (2,1,0,-1). $n=-4$ can be used for millicharged dark matter \cite{mcharge1,mcharge2,Aboubrahim:2021ohe,PhysRevD.98.103005,PhysRevLett.121.011102,PhysRevD.98.023013} as in this case the scattering of baryons takes place via a Coulomb-like potential. Moreover, $n=-4$ is applicable in several such cases, namely hadronically interacting DM, millicharge DM, the Baryon Acoustic Oscillations (BAO) signal etc. In the present work, we choose $n=-4$, which is also considered in several similar recent treatments (e.g. Ref.~\cite{munoz,upala,21cm_jan,21cm_feb,21cm_mar,21cm_upala,Mahdawi_2018,rennan_3GeV}). As a result the baryon-dark matter cross-section term takes the form 
	\begin{equation}
		\sigma = \left(\sigma_0\right) v^{-4}= \left(\sigma_{41} \times 10^{-41}\,{\rm cm^{2}}\right) v^{-4},
	\end{equation}
	where $\sigma_{41}$ is a dimensionless quantity defined as $\sigma_{41}=\sigma_0/\left(10^{-41}{\rm cm^{2}}\right)$.
	
	We  assume  the value of $\sigma_0$ to lie within $10^{-42} \sim 10^{-40}\rm{cm^2}$, which corresponds to the scalar cross-section bound obtained from the recent  experiments on direct dark matter detection (obtained by extrapolating the permissible zone for the dark matter range $0.1$ GeV$\leq m_{\chi} \leq 3$ GeV from recent experiments \cite{xenon1t,lux,pandax2}). Now, for $n=-4$, the evolution equation of $V_{\chi b}$ can be written as,
	\begin{equation}
		\dfrac{{\rm d}V_{\chi b}}{{\rm d}t} = -D(V_{\chi b}) = \dfrac{\rho_m \sigma_0}{m_b+m_{\chi}}\dfrac{1}{V_{\chi b}^2}F(r),
		\label{eq:dvdt}
	\end{equation}
	where $D(V_{\chi b})$ is the drag term.
	In the above expression, $m_b$ and $m_{\chi}$ are the average mass of baryon and DM particles respectively, $\rho_m$ is the matter density and the function $F(r)$ is given by
	\begin{equation}
		F(r)={\rm erf}(r/\sqrt{2})-\sqrt{2/\pi} r e^{-r^2/2},
	\end{equation}
	where $r=V_{\chi b}/u_{th}$, $u_{th}=\sqrt{T_b/m_b+T_{\chi}/m_{\chi}}$ ($T_b$ and $T_{\chi}$ are the temperature of baryon and dark matter fluid respectively).
	So, the evolution equation of $V_{\chi b}$ is given by
	\begin{equation}
		\dfrac{{\rm d} V_{\chi b}}{{\rm d}z}=\dfrac{V_{\chi b}}{1+z}+\dfrac{D(V_{\chi b})}{(1+z)H(z)}
		\label{eq:V_chib}
	\end{equation}
	Now from the work of Mu\~{n}oz \emph{et~al.} \cite{munoz}, it can be shown that, the heating rate of the baryonic fluid is given by
	\begin{equation}
		\dfrac{{\rm d} Q_b}{{\rm d} t}=\dfrac{2 m_b \rho_{\chi}\sigma_0 e^{-r^{2}/2} (T_{\chi}-T_b)}{(m_b+m_{\chi})^2\sqrt{(2\pi)} u_{th}^3}+\dfrac{\rho_{\chi}}{\rho_b+\rho_{\chi}}\dfrac{m_{\chi} m_b}{m_{\chi}+m_b} V_{\chi b} D(V_{\chi b}).
		\label{eq:dqbdt}
	\end{equation}
	Similarly, the heating rate of DM fluid can be formulated as
	\begin{equation}
		\dfrac{{\rm d} Q_{\chi}}{{\rm d} t}=\dfrac{2 m_{\chi} \rho_{b}\sigma_0 e^{-r^{2}/2} (T_b-T_{\chi})}{(m_b+m_{\chi})^2\sqrt{(2\pi)} u_{th}^3}+\dfrac{\rho_{b}}{\rho_b+\rho_{\chi}}\dfrac{m_{\chi} m_b}{m_{\chi}+m_b} V_{\chi b} D(V_{\chi b}).
		\label{eq:dqchidt}
	\end{equation}

	\section{\label{sec:T_evol} Thermal Evolution}
	
	In order to investigate the effects of viscous dark energy, primordial black holes and all possible interactions of dark matter, {\it viz.}, baryon-dark matter interaction, dark matter annihilation and dark matter decay on the 21-cm signal, we have to
	first formulate the thermal evolution of the baryons ($T_b$) and dark matter ($T_{\chi}$)  as a function of redshift $z$, given by \cite{corr_equs,munoz,BH_21cm_1,BH_21cm_2,rupadi,21cm_mar,21cm_jan}
	\begin{equation}
		(1+z)\dfrac{{\rm d} T_\chi}{{\rm d} z} = 2 T_\chi - \dfrac{2}{3 H(z)}\dfrac{{\rm d} Q_{\chi}}{{\rm d} t}-\dfrac{\Omega_{\chi}}{\Omega_m}\dfrac{2}{3 H(z)}\dfrac{m_{\chi}}{\rho_{\chi}}\dfrac{{\rm d}Q}{{\rm d}V{\rm d}t}, 
		\label{eq:T_chi}
	\end{equation}
	\begin{equation}
		(1+z)\frac{{\rm d} T_b}{{\rm d} z} = 2 T_b + \frac{\Gamma_c}{H(z)}
		(T_b - T_\gamma)-\frac{2}{ 3 H(z)}\dfrac{{\rm d} Q_{b}}{{\rm d} t}-\frac{2}{3 k_B H(z)}
		\frac{K_{\rm heat}}{1+f_{\rm He}+x_e}-\dfrac{\Omega_{b}}{\Omega_m}\dfrac{2}{3 H(z)}\dfrac{m_{b}}{\rho_{b}}\dfrac{{\rm d}Q}{{\rm d}V{\rm d}t}. 
		\label{eq:T_b}
	\end{equation}
	In the above expressions, $x_e$ and $f_{\rm He}$ are the ionization fraction and the fractional abundance of Helium respectively, $T_{\gamma}$ ($=2.725(1+z) K$) is the background temperature, $K_{\rm heat}$ corresponds to baryon heating, and $H(z)$ is the Hubble constant, which varies with $z$ according to the viscous dark energy models (see \autoref{eq:mod1}, \ref{eq:mod2} and \ref{eq:mod3}). In \autoref{eq:T_chi} and \autoref{eq:T_b}, the last terms of each expression correspond to the dark matter and baryon heating respectively due to the entropy produced as an outcome of viscous flow of DE. The first terms of both those equations describe the same for Hubble expansion. The second term of \autoref{eq:T_chi} denotes the contribution of baryon-DM interaction in the heating/cooling of the dark matter and the third term of \autoref{eq:T_b} is the same in the case of baryons. The effects of the Compton scattering ($\Gamma_c$ is the Compton interaction rate) and the combined heating due to Hawking radiation, DM annihilation and DM decay are included in the second and fourth terms of \autoref{eq:T_b}, respectively ($K_{\rm heat}$ is described later in \autoref{KBH}).
	
	The evolution equation of the ionization fraction $x_e$ ($=n_e/n_H$, where $n_e$ and $n_H$ are the free electron number density and hydrogen number density respectively) depends on the model dependent Hubble parameter $H$ alongside $T_{\gamma}$ and $T_b$, given by \cite{munoz, BH_21cm_2,21cm_mar,PhysRevD.98.023501,21cm_jan,rupadi},	
	\begin{equation}
		\frac{{\rm d} x_e}{{\rm d} z} = \frac{1}{(1+z)\,H(z)}\left[I_{\rm Re}(z)- I_{\rm Ion}(z)-I_{\rm heat}(z)\right].
		\label{eq:xe}
	\end{equation}
	In this expression, $I_{\rm Ion}(z)$ and $I_{\rm Re}(z)$ are the standard ionization rate and the standard recombination rate respectively \cite{yacine,hyrec11,munoz}, the
	difference of which can be expressed as \cite{peeble,hyrec11,hummer,pequignot,seager,pequignot,BH_21cm_5,21cm_feb},
	\begin{equation}
		I_{\rm Re}(z)-I_{\rm Ion}(z) = C_P\left(n_H \alpha_B x_e^2-4(1-x_e)\beta_B 
		e^{-\frac{3 E_0}{4 k_B T_{\gamma}}}\right),
		\label{eq:xe_comp}
	\end{equation}
	where $C_P$ is the Peebles-C factor \cite{peeble,hyrec11}, $\alpha_B$ and $\beta_B$ are the case B recombination and ionization coefficients respectively \cite{pequignot,seager,BH_21cm_5,21cm_jan} and $E_0=13.6$ eV.
	
	The fraction of total energy deposited from Hawking radiation, DM annihilation and DM decay in the form of heat and ionization are respectively $\chi_h=(1+2 x_e)/3$ and $\chi_i=(1-x_e)/3$ \cite{BH_21cm_2,chen,BH_21cm_4,PhysRevD.76.061301,Furlanetto:2006wp}. Consequently, the contribution of those sources in the form of baryon heating ($K_{\rm heat}$, see \autoref{eq:T_b}) and hydrogen ionization ($I_{\rm heat}$, see \autoref{eq:xe}) can be expressed as follows (for reference see \cite{PhysRevD.98.023501,21cm_mar,21cm_jan,rupadi}),
	\begin{equation}
		K_{\rm heat}=\chi_{h} \frac{1}{n_b} \left(f_{\rm BH}(z)\left.\dfrac{{\rm d} E}{{\rm d}V {\rm d}t}\right|_{\rm{BH}}+f_{\rm ann.}(z)\left.\dfrac{{\rm d} E}{{\rm d}V {\rm d}t}\right|_{\rm{ann.}}+f_{\rm dec.}(z)\left.\dfrac{{\rm d} E}{{\rm d}V {\rm d}t}\right|_{\rm{dec.}}\right), \label{KBH}
	\end{equation}
	\begin{equation}
		I_{\rm heat}=\chi_{i} \frac{1}{n_b} \frac{1}{E_0}\left(f_{\rm BH}(z)\left.\dfrac{{\rm d} E}{{\rm d}V {\rm d}t}\right|_{\rm{BH}}+f_{\rm ann.}(z)\left.\dfrac{{\rm d} E}{{\rm d}V {\rm d}t}\right|_{\rm{ann.}}+f_{\rm dec.}(z)\left.\dfrac{{\rm d} E}{{\rm d}V {\rm d}t}\right|_{\rm{dec.}}\right). \label{IBH}
	\end{equation}
	In the above expressions, the factors $f_{\rm BH}(z)$, $f_{\rm ann.}(z)$ and $f_{\rm dec.}(z)$ are the total fraction of the injected energy deposited into the IGM at redshift $z$ due to PBH evaporation, dark matter annihilation and dark matter decay, respectively \cite{corr_equs,fcz001,fcz002,fcz003,fcz004}. In the present analysis, $f_{\rm ann.}(z)$ and $f_{\rm dec.}(z)$ are calculated by considering only $e^+ e^-$ and photon final state product from DM annihilation and decay  \cite{rupadi,BH_21cm_1,corr_equs,fcz001,fcz002,fcz003}. 

	\section{\label{sec:result} Combined effect on the 21-cm brightness temperature}
	
	In the present work, our aim is to investigate the effect of viscous dark energy in the global 21-cm scenario, and thus estimate the bounds on the viscous dark energy model parameters.	
	As we have seen from the above analysis, the viscous flow of DE can modify the Hubble parameter significantly for some ranges of the model parameters. As a result both baryon and dark matter are cooled down comparatively faster. Besides this cooling effect, entropy is also produced due to viscous flow, which heats up both baryons and dark matter components. We incorporate various dark matter phenomena, {\it viz.}, baryon heating/cooling effects by the baryon-DM interaction, dark matter annihilation and dark matter decay, Compton scattering and PBH evaporation in our analysis for calculating the 21-cm brightness temperature. 
	

	The variation of spin temperature $T_s$, the radiation temperature $T_{\gamma}$, and hence the 21-cm brightness temperature $T_{21}$ are estimated by computing the baryon and dark matter temperatures at different $z$, which are obtained by solving several coupled equations  (\autoref{eq:PBH}, \ref{eq:V_chib}, \ref{eq:T_chi}, \ref{eq:T_b}, and \ref{eq:xe}) simultaneously. It is assumed that, initially at $z\approxeq1010$, baryons and radiation were tightly coupled. So, at that epoch the baryon temperature ($T_b$) and the background temperature ($T_{\gamma}$) were exactly equal. The initial temperature of the dark matter fluid is negligible at $z=1010$, as the DM particles are assumed to be in the form
	of cold dark matter (if a slightly warm dark matter is taken into account, the thermal evolution remains almost unaffected \cite{munoz}). The initial relative velocity ($V_{\chi b}$) is considered to be $\sim$29 km/s \cite{munoz,rel_vel_1st,PhysRevD.89.083506}.
	
	\begin{figure*}
		\centering
		\begin{tabular}{cc}
			\includegraphics[width=0.5\textwidth]{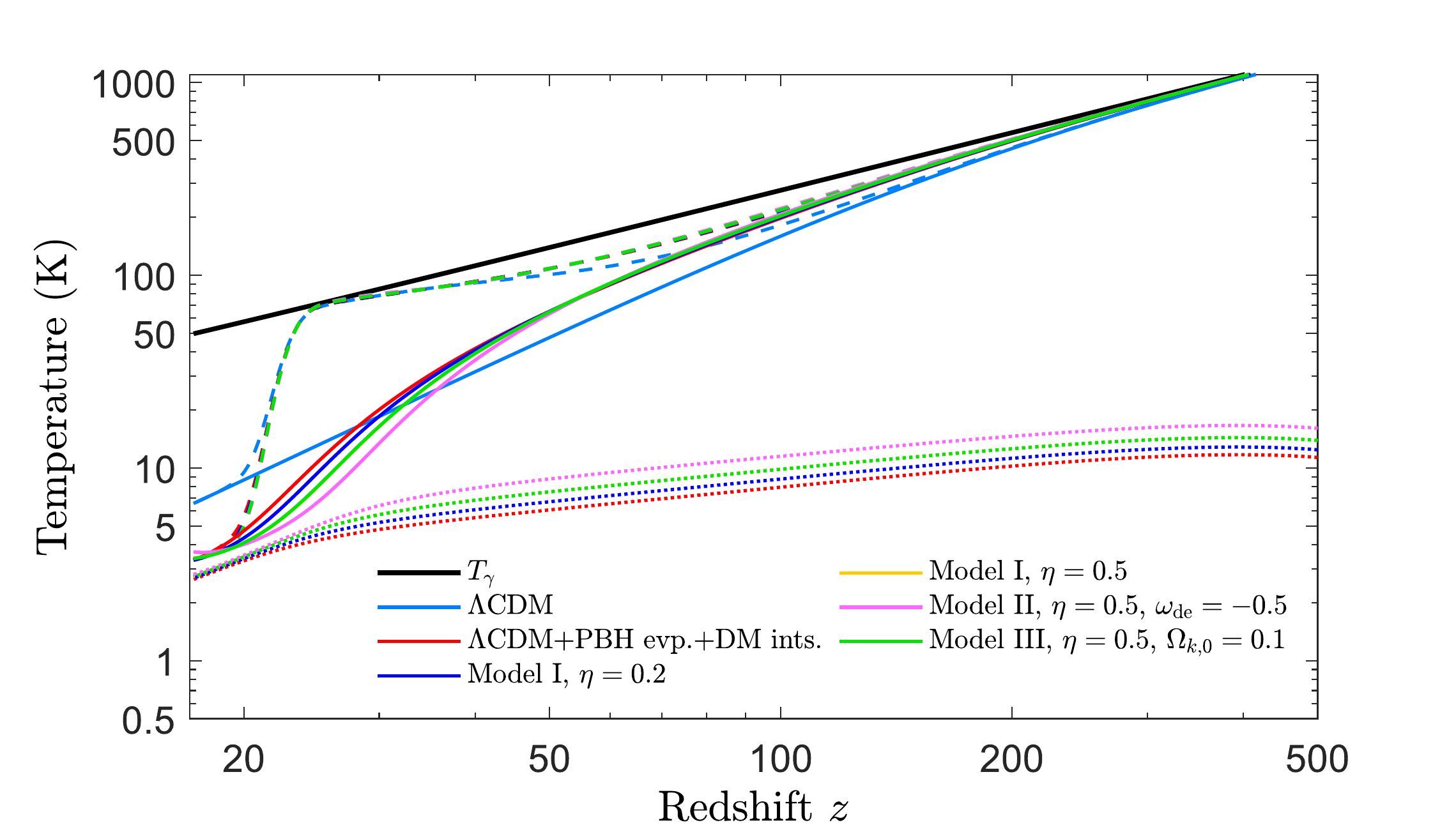}&
			\includegraphics[width=0.5\textwidth]{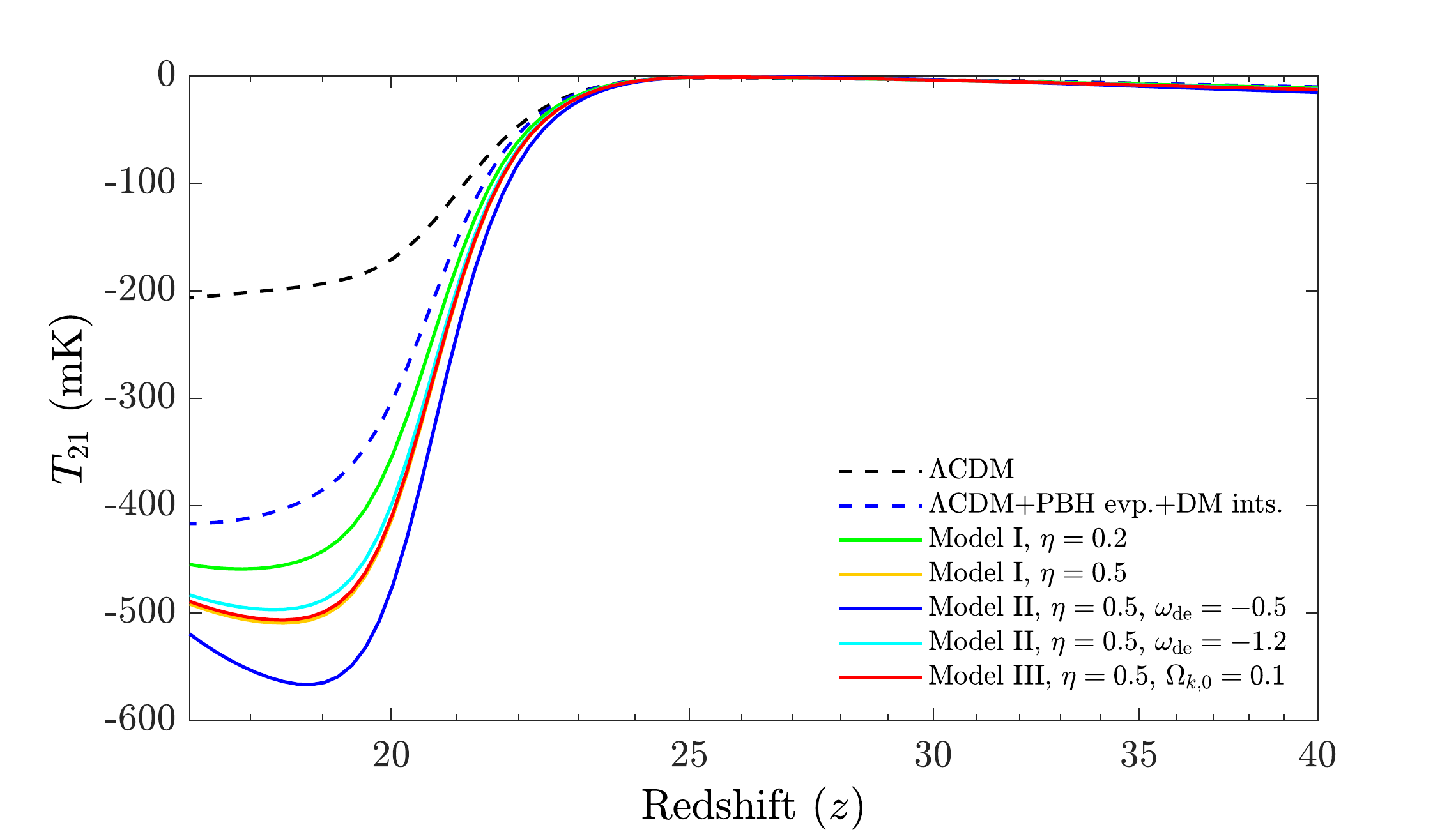}\\
			(a)&(b)\\
		\end{tabular}
		\caption{\label{fig:tspin_t21} (a) Variations of dark matter temperature $T_{\chi}$, baryon temperature $T_b$ and the corresponding spin temperature $T_s$ with redshift $z$ for different chosen values of VDE models and parameters. (b) Variations of  brightness temperature with redshift $z$ for different chosen values of VDE models and parameters. In individual cases of \autoref{fig:tspin_t21}(a) and \autoref{fig:tspin_t21}(b), dark matter particle mass $m_{\chi}=1$ GeV, $\sigma_{41}=1$, DM decay lifetime $\tau_{\chi}=10^{-28}$ s PBH mass $M_{\rm BH}=1.5\times 10^{14}$ g and initial mass fraction of PBH $\beta_{\rm BH}=10^{-29}$ are chosen. $\langle \sigma v \rangle$ for individual cases are calculated as described in \autoref{sec:DM_ann_dec}. See text for further details.}
	\end{figure*}
	
	In \autoref{fig:tspin_t21}(a), the evolution of baryon temperature $T_b$ (solid lines) for different VDE models and model parameters are shown by incorporating the above mentioned modification in Hubble parameter and the additional entropy produced due to viscous flow of dark energy. The cyan lines represent the case of the $\Lambda$CDM model while the red lines denote the same where the effect of PBH evaporation as well as baryon-DM interaction, dark matter annihilation and dark matter decay are included. The corresponding spin temperatures $T_s$ and dark matter temperatures for all the cases are plotted in the same graph using  dotted lines and  dashed lines of same colours, respectively. Here the solid black line shows the variation of $T_{\gamma}$ with $z$. The variations of $T_b$, $T_{\chi}$ and $T_s$ in presence of different VDE models are shown using different colour. From this figure (\autoref{fig:tspin_t21}(a)) it can be seen that at the higher redshifted region, the spin temperatures of different cases almost overlap with each other. However, at the end of cosmic dark age ($z\lessapprox 25$), the lines of $T_s$ start falling and begin to overlap with the corresponding baryon temperatures ($T_b$) at $z\lessapprox20$ (this cosmic phenomenon is known as the Wouthuysen-Field effect). 
	
	Similar variations of brightness temperature ($T_{21}$) are represented graphically in \autoref{fig:tspin_t21}(b). In this figure, the black dashed line describes  $T_{21}$ for the $\Lambda$CDM model, which gives $T_{21}\approx -200$ mK at $z=17.2$. The blue dashed line denotes the same where the effect of primordial black hole evaporation, baryon-DM interaction, DM annihilation and DM decay are included. The other colour lines of \autoref{fig:tspin_t21} represent the brightness temperature for different VDE models and different values of $\eta$. It can be seen that as the cooling due to baryon-DM interaction is incorporated in the calculation, $T_{21}$ falls significantly (comparing the blue and black dashed line of \autoref{fig:tspin_t21}(b)). In presence of viscosity the brightness temperature falls further, which indicates that the resultant effect of the viscous dark energy models essentially cools down the baryonic fluid in spite of producing entropy due to viscous flow. It is also observed that the effect of the VDE Model III, {\it i.e.}, the effect of curvature is small in comparison with that of the viscosity parameter $\eta$ and the DE equation of state parameter $\omega_{\rm de}$.  It is to be mentioned that all plots of \autoref{fig:tspin_t21}(a) and \autoref{fig:tspin_t21}(b) are plotted for the values of dark matter mass $m_{\chi}=1$ GeV, $\sigma_{41}=1$, PBH mass $M_{\rm BH}=1.5\times 10^{14}$ g, initial mass fraction of PBH, $\beta_{\rm BH}=10^{-29}$, $\tau_{\chi}=10^{24}s$ and the value of $\langle \sigma v \rangle$ is chosen as described in \autoref{sec:DM_ann_dec}.
	
	\begin{figure*}
		\centering
		\includegraphics[trim=0 40 0 0, clip,width=\textwidth]{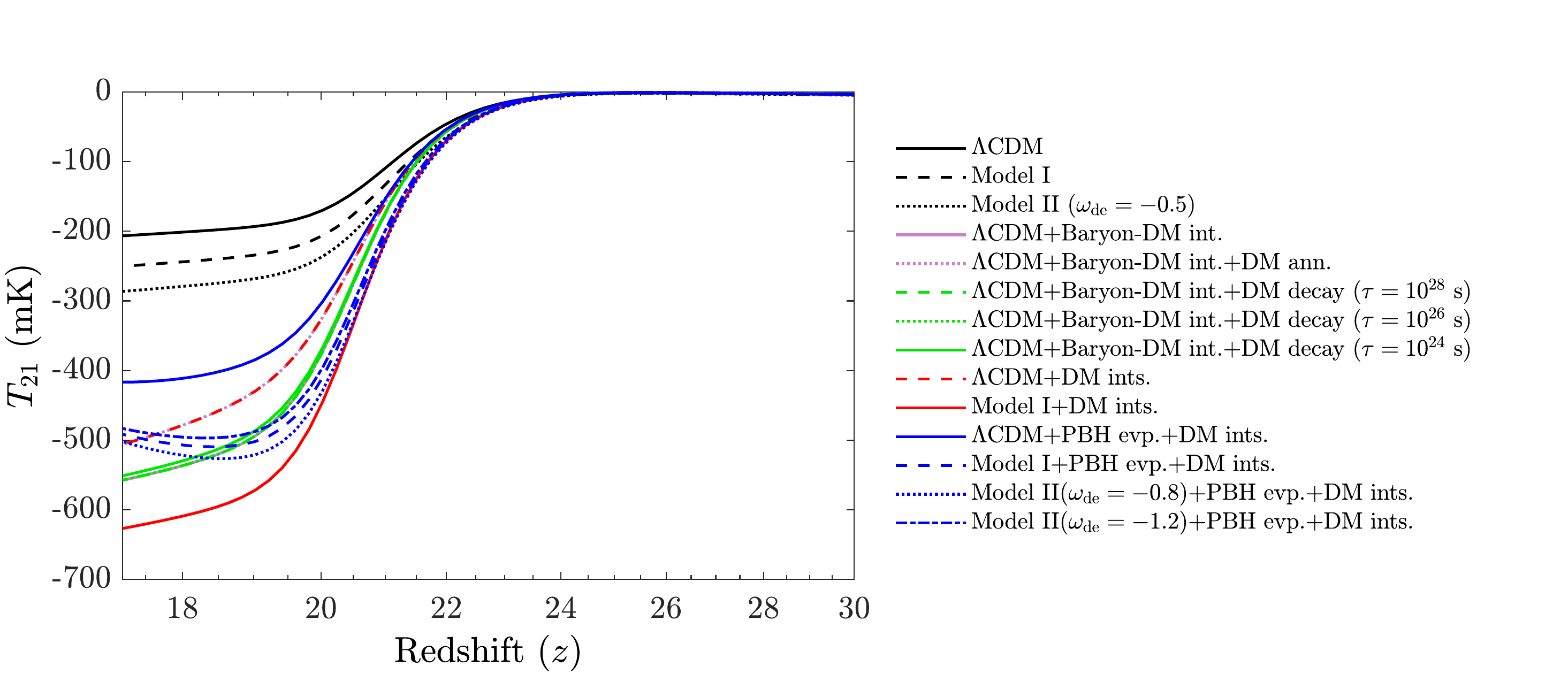}
		\caption{\label{fig:tspin_t21_int} Variation of $T_{21}$ with redshift $z$ for different cases. See text for details.}
	\end{figure*}
	
	The comparative influence of the various physical factors (viscosity, baryon-DM scattering, velocity averaged DM annihilation cross-section, DM decay lifetime and PBH decay) on the brightness temperature $T_{21}$ is elucidated in the \autoref{fig:tspin_t21_int}. Here all the curves of $T_{21}$ with VDE models are plotted with $\eta=0.5$. In this figure, the black solid line shows the variation of $T_{21}$ for the $\Lambda$CDM model, while the black dashed line denotes the same for the VDE Model I in absence of the contribution of any DM interaction and PBH evaporation. From this, one can observe that unlike baryon cooling via baryon-DM scattering, the viscosity cannot provide the required cooling alone in order to sustain the EDGES consequence ($T_{21}<-300$ mK). However, the VDE Model II with $\Omega_{\rm de}=-0.5$ provides larger than expected cooling due to the additional modification in the Hubble parameter in the form of $\omega_{\rm de}$ (black dotted line). The difference between the $T_{21}$ for the cases of $\Lambda$CDM model and VDE Model I increases remarkably as the cooling due to baryon-DM interaction is incorporated along with the heating due to the phenomena of DM annihilation and DM decay (red dashed line and red solid line). However, the incorporation of the heating effect due to the Hawking radiation from PBHs slightly decreases this separation. The cooling effect by the viscous flow of dark energy further enhances in presence of higher values of $\omega_{\rm de}$ (blue dotted line and blue dash-dotted line).

	From \autoref{fig:tspin_t21_int}, the heating effects due to PBH evaporation, DM annihilation and DM decay on the global 21-cm signal can be observed. The heating in the form of Hawking radiation is clearly seen by comparing the red dashed line and blue solid line for the $\Lambda$CDM model. A similar effect in the VDE Model I is observed by comparing the solid red line and the blue dashed line. In this plot the solid violet  line almost overlaps
	with the green dashed line and the green dotted line. From the green dashed,  dotted  and  solid lines it appears that the heating due to DM decay  does not manifest any significant footprint on the temperature. However, the violet dotted line (overlapped by the red dashed line), which represents the case of DM annihilation in addition to the baryon-DM scattering, shows a significant higher temperature in comparison to the same for baryon-DM scattering alone (solid violet line).

\begin{figure}
		\centering
		\includegraphics[width=0.6\linewidth]{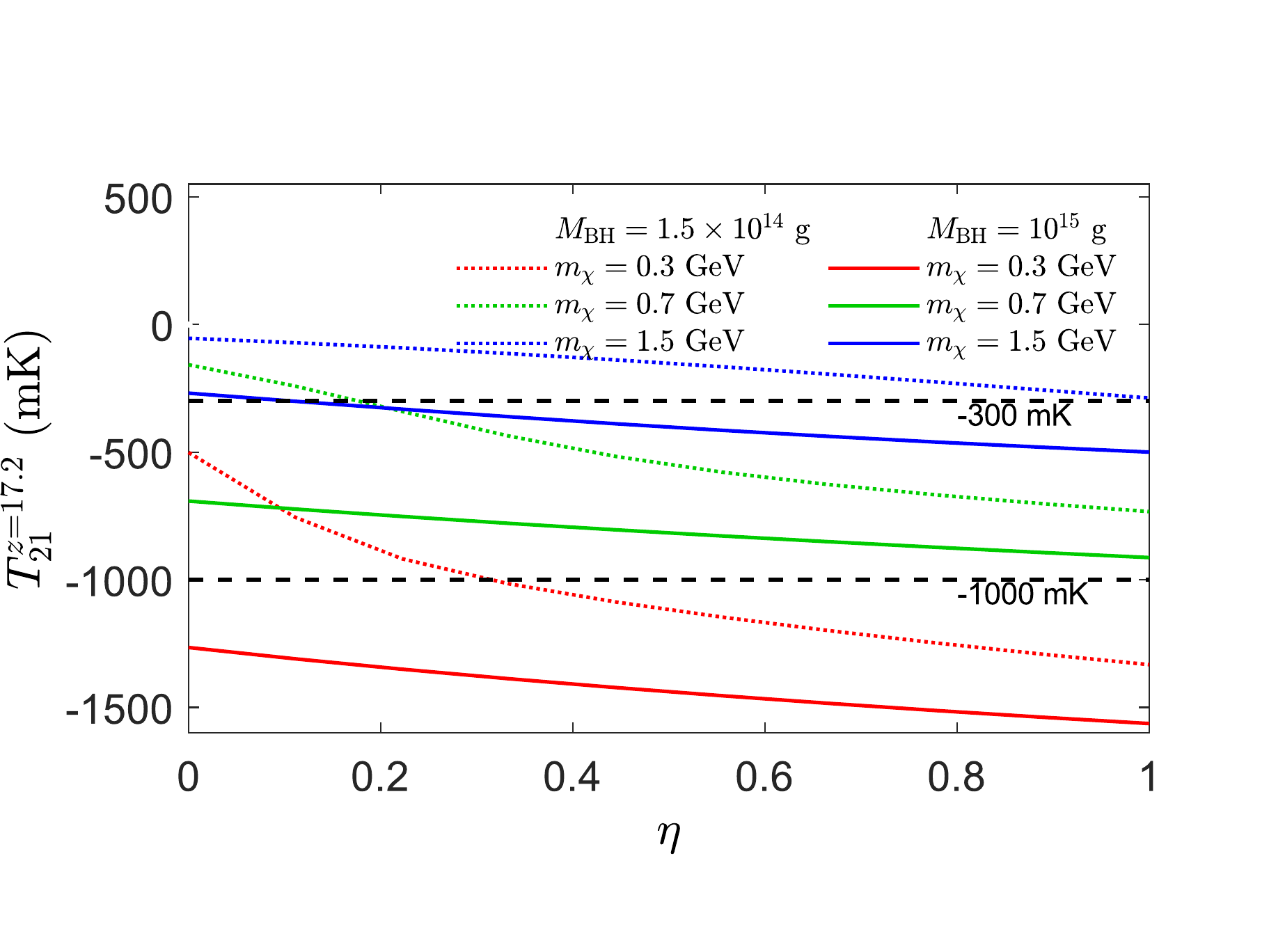}
		\caption{\label{fig:mod1_t21} Variation of $T_{21}^{z=17.2}$ with $\eta$ for different chosen combinations of PBH mass $M_{\rm BH}$ and dark matter mass $m_{\chi}$. See text for details.}
	\end{figure}

	We now  extend our analysis to constrain the model parameters that can account for the EDGES observation of  $-300\geq T_{21}\geq -1000$ at $z=17.2$. To this end we define a new parameter $T_{21}^{z=17.2}$ which measures the brightness temperature at $z=17.2$. In \autoref{fig:mod1_t21} we display the variation of $T_{21}^{z=17.2}$ with the viscosity parameter $\eta$ for the VDE Model I, where different combinations of dark matter mass $m_{\chi}$ and primordial black hole mass $M_{\rm BH}$ considered. During this analysis, we fix the initial mass fraction of PBH at $\beta=10^{-29}$, DM decay lifetime $\tau_{\chi}=10^{28}$s, baryon-DM scattering parameter $\sigma_{41}=1$ and the velocity averaged DM annihilation cross-section is estimated as described in \autoref{sec:DM_ann_dec}. From this figure it can be seen that for each combination of $m_{\chi}$ and $M_{\rm BH}$, the value of $T_{21}^{z=17.2}$ falls with increasing $\eta$ and the corresponding slope is in some cases larger for lower $\eta$. This result clearly establishes the cooling of the 21-cm brightness temperature in presence of viscosity in the dark energy fluid. Moreover, it is also observed that lower mass of the dark matter particle $m_{\chi}$ provides lower values of $T_{21}^{z=17.2}$. The reason behind this larger cooling by lower $m_{\chi}$ is that for lower values of $m_{\chi}$ the number density of DM particles ({\it i.e.}, $n_{\chi}$) is high (since the abundance of dark matter is fixed). As a result, a larger number of DM particles interact with baryons (keeping baryon-DM interaction cross-section fixed) and cool down the baryons faster. This result agrees with several recent results in other treatments \cite{munoz,21cm_jan,21cm_feb,21cm_mar,PhysRevD.98.023013,PhysRevLett.121.011102,PhysRevD.98.103005,21cm_upala,rupadi}. In addition to the above effects, one can see that as PBHs having larger mass radiate a lower amount of energy,  the lines correspond to $M_{\rm BH}= 10^{15}$ g are cooler than those for $M_{\rm BH}=1.5 \times 10^{14}$ g .
	
\begin{figure*}
		\centering
		\begin{tabular}{cc}
			\includegraphics[trim=0 40 80 60, clip, width=0.48\textwidth]{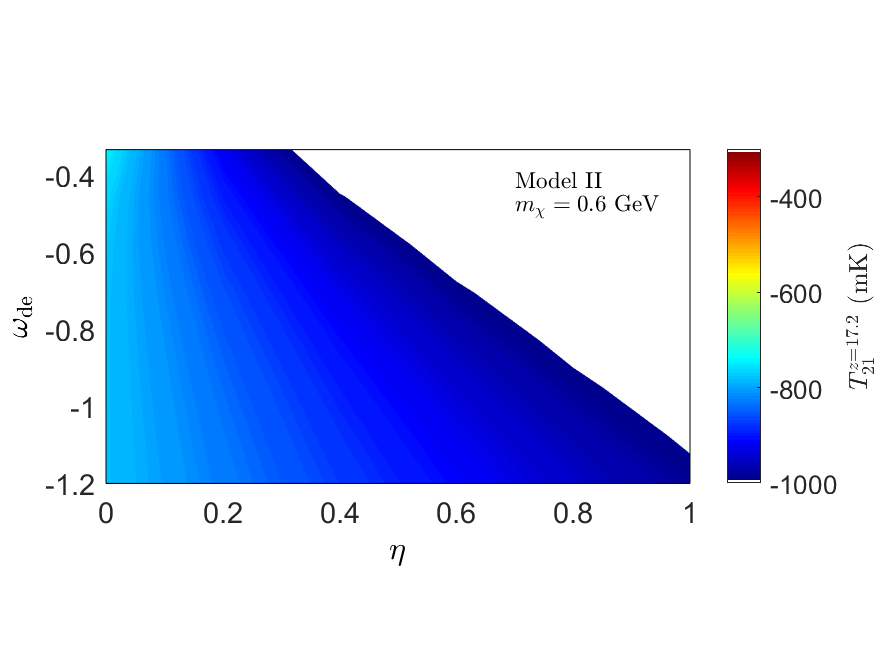}&
			\includegraphics[trim=0 40 80 60, clip, width=0.48\textwidth]{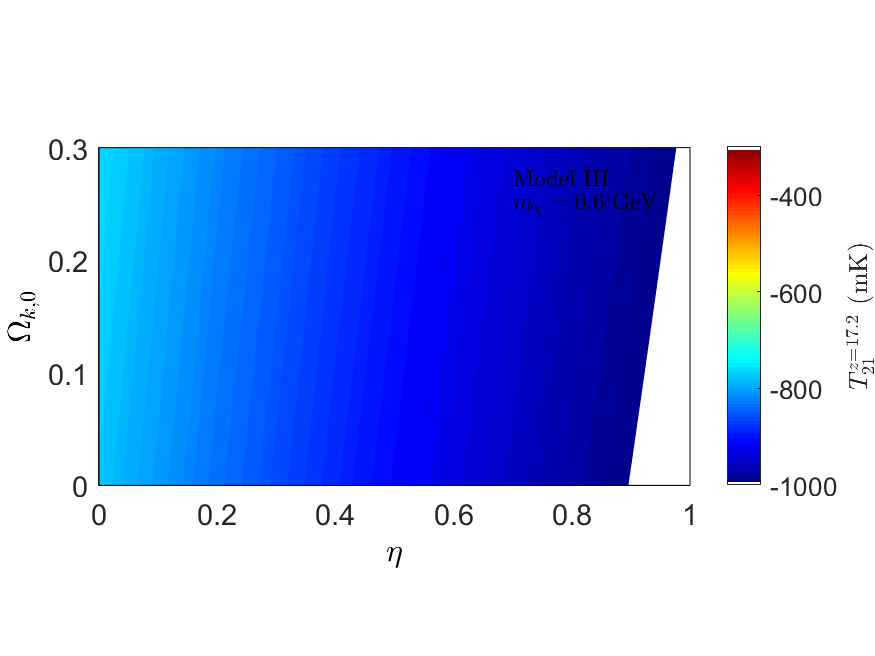}\\
			(a)&(b)\\
			\includegraphics[trim=0 40 80 60, clip, width=0.48\textwidth]{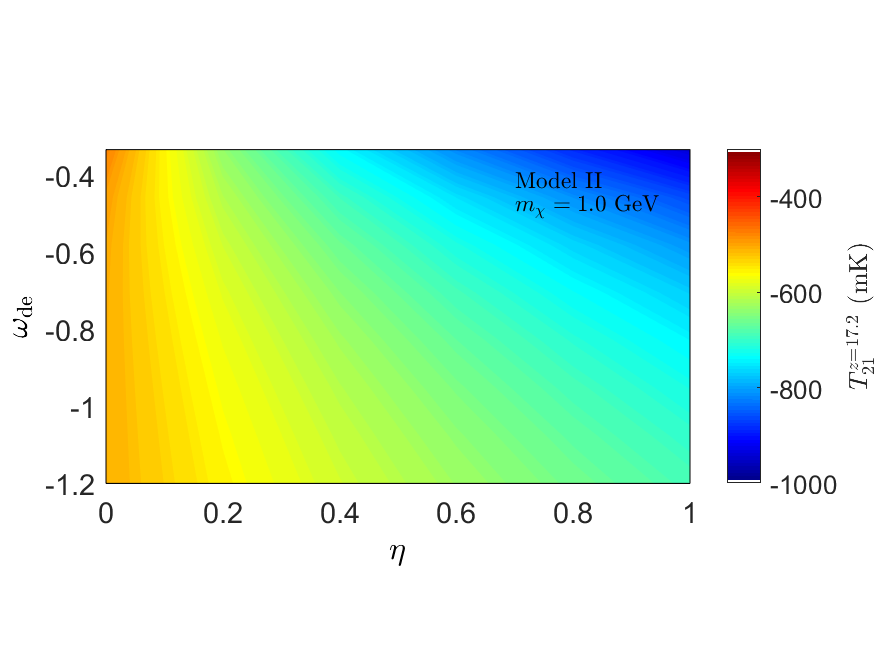}&
			\includegraphics[trim=0 40 80 60, clip, width=0.48\textwidth]{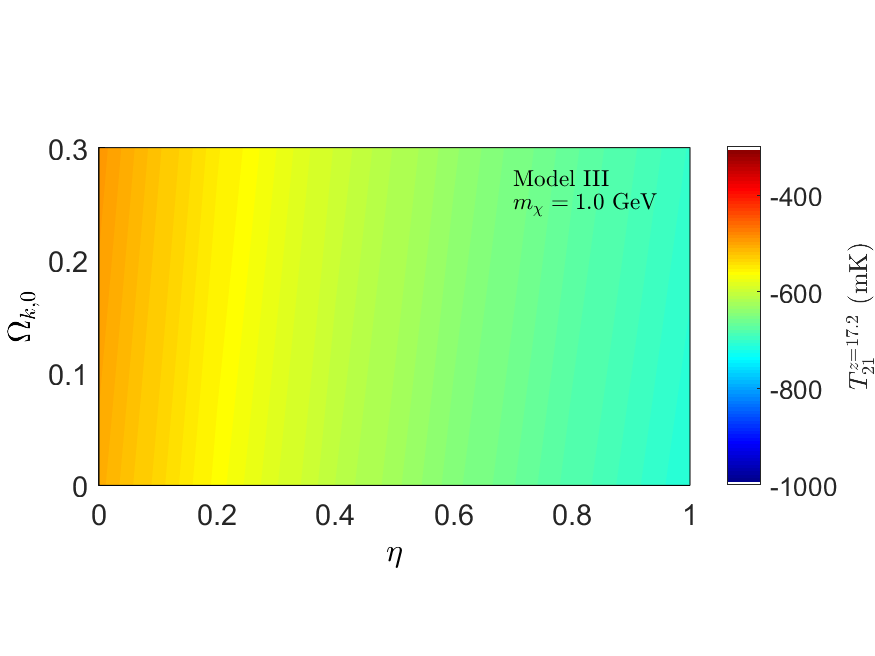}\\
			(c)&(d)\\
			\includegraphics[trim=0 40 80 60, clip, width=0.48\textwidth]{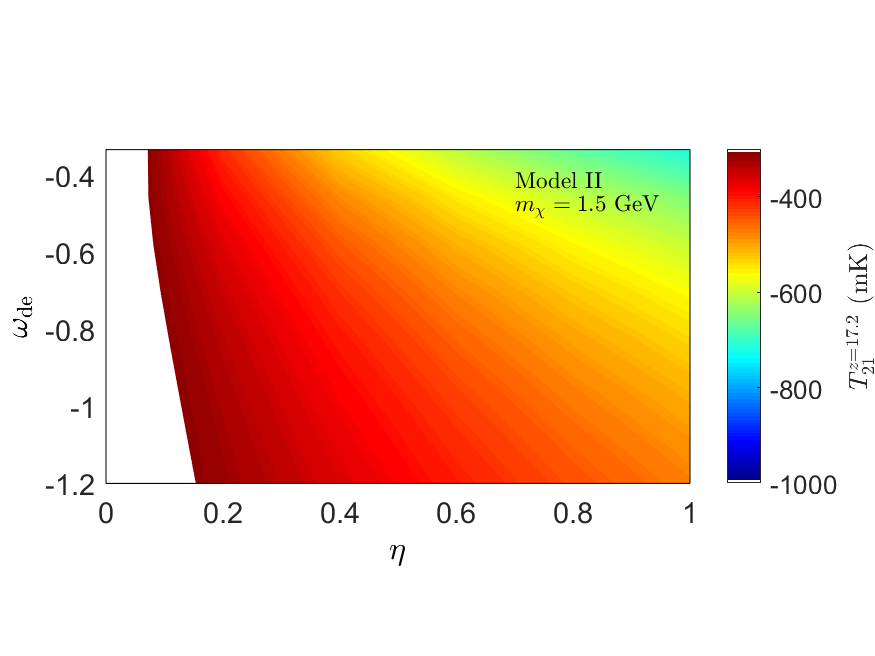}&
			\includegraphics[trim=0 40 80 60, clip, width=0.48\textwidth]{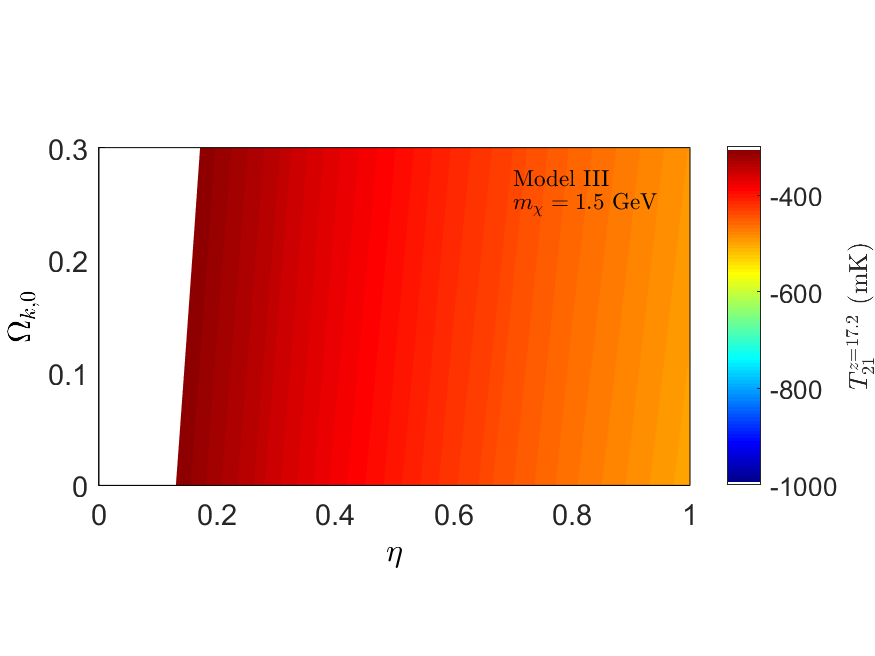}\\
			(e)&(f)\\
		\end{tabular}
		\begin{tabular}{c}
			\includegraphics[trim=0 10 0 220,clip, width=0.9\textwidth]{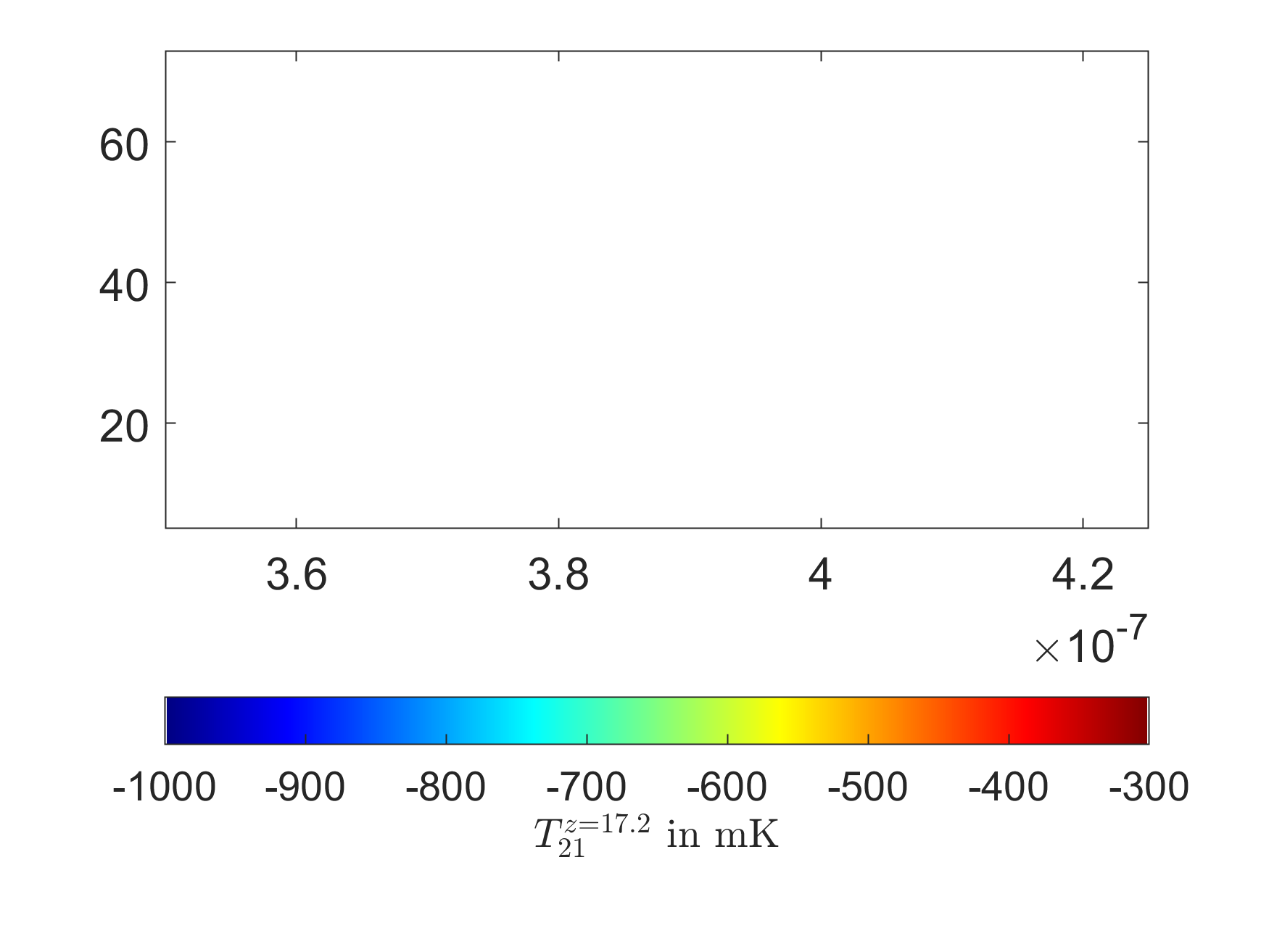}\\
		\end{tabular}
		\caption{\label{fig:mod2_3_cont} Variation of $T_{21}^{z=17.2}$ with VDE model parameters for Model II (left side plots) and Model III (right side plots). Here \autoref{fig:mod2_3_cont}(a) and  \autoref{fig:mod2_3_cont}(b) are calculated for $m_{\chi}=0.6$ GeV, while \autoref{fig:mod2_3_cont}(c), \autoref{fig:mod2_3_cont}(d) are for $m_{\chi}=1.0$ GeV and \autoref{fig:mod2_3_cont}(e), \autoref{fig:mod2_3_cont}(f) for $m_{\chi}=1.5$ GeV.}
	\end{figure*}

	Next, similar analyses are carried out for both the VDE Models II and III. In \autoref{fig:mod2_3_cont}(a), (c) and (e) the contour plots are furnished for $m_{\chi}=0.6$ GeV, 1.0 GeV and 1.5 GeV respectively, where the VDE Model II is considered, along with primordial black hole mass $M_{\rm BH}=10^{15}$ g, $\beta_{\rm BH}=10^{-29}$, $\tau_{\chi}=10^{28}$s. Similar contour plots are displayed in \autoref{fig:mod2_3_cont}(b), (d) and (f) respectively, for the case of the VDE Model III. In all plots of \autoref{fig:mod2_3_cont}, different colours represent different values of $T_{21}^{z=17.2}$ in mK, which are mentioned in the colourbar furnished at the bottom of \autoref{fig:mod2_3_cont}. The white regions in \autoref{fig:mod2_3_cont}(a), (e) and (f) indicate the areas beyond the allowed zone according to the EDGES result. From \autoref{fig:mod2_3_cont}(a), (c) and (e) it can be seen that higher values of $\omega_{\rm de}$ essentially provide lower $T_{21}^{z=17.2}$, as also observed in the \autoref{fig:tspin_t21}(b). However, at lower viscosity ($\eta \lessapprox 0.18$) and higher $\omega_{\rm de}$ ($\omega_{\rm de}\gtrapprox-0.5$), a small inverse variation is observed ({\it i.e.} $T_{21}^{z=17.2}$ increases with $\omega_{\rm de}$, see top-left region of \autoref{fig:mod2_3_cont}(a), (c) and (e)). 
	
\begin{figure*}
		\centering
		\includegraphics[width=0.6\linewidth]{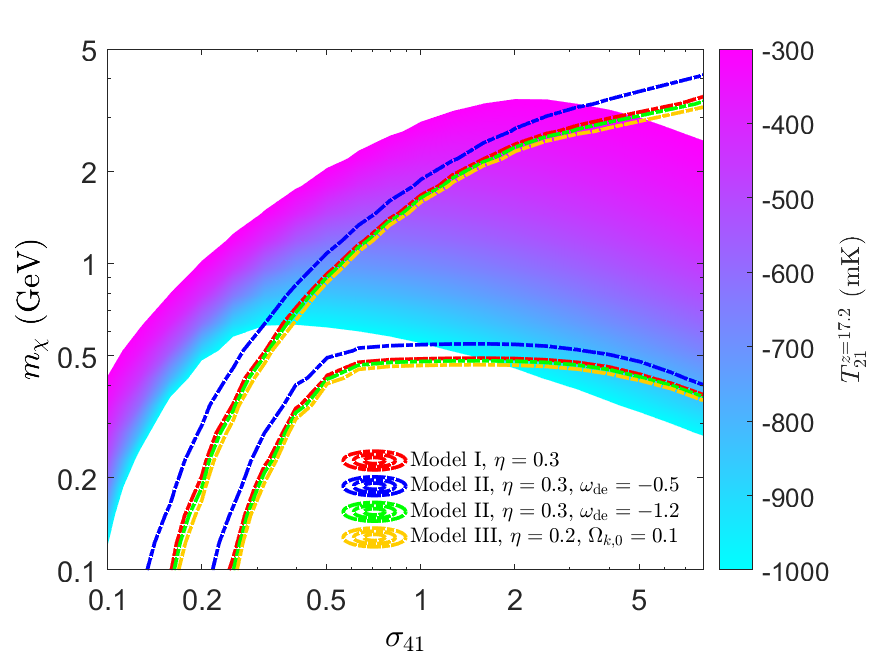}
		\caption{\label{fig:mchi_sigma} The allowed region in the $m_{\chi}$ - $\sigma_{41}$ space. The coloured shaded region represents the allowed region for the case of $\Lambda$CDM model in absence of the viscosity in the dark energy fluid, while the area between the coloured lines of same colour represent the similar bounds for different VDE models. See text for details.}
	\end{figure*}

	It is also observed that the variation of $T_{21}^{z=17.2}$ with $\omega_{\rm de}$ increases when higher values of $\eta$ are taken into account in case of the VDE Model II. On the other hand, in case of the Model III, higher values of curvature provide higher values of $T_{21}^{z=17.2}$ (see \autoref{fig:mod2_3_cont}(b), (d) and (f)), but the variation of $T_{21}^{z=17.2}$ with $\Omega_{k,0}$ is small in comparison to that with $\eta$.  We find that for $m_{\chi}=0.6$ GeV, the region for $\eta \leq 1$ lies within the allowed region according to the EDGES result. However in presence of higher mass DM particles the brightness temperature of individual points increases. If further higher values of $m_{\chi}$ are 
	considered, the brightness temperature of the 21-cm spectra at lower values of $\eta$ lies beyond the allowed region. From \autoref{fig:mod2_3_cont}(e) and (f) (for VDE Model II and III respectively) one can see that in presence of such higher DM masses the brightness temperature may lie within the EDGES allowed region, if significant bulk viscosity of dark energy is considered \cite{rennan_3GeV,21cm_jan,21cm_feb}. A detailed analysis of this phenomena is presented in \autoref{fig:mchi_sigma}.
	
	Finally, we estimate the allowed region in the $m_{\chi}$-$\sigma_{41}$ plane for different VDE models. In \autoref{fig:mchi_sigma}, the coloured region represents the allowed region in the $m_{\chi}$-$\sigma_{41}$ plane, where the effect of viscous dark energy, PBH evaporation, dark matter annihilation and decay are not incorporated. In this plot, different colours of the coloured region correspond to different values of $T_{21}^{z=17.2}$, as described in the colourbar of the same figure. The presence of viscous flow of dark energy along with the combined effect of Hawking radiation from PBHs, baryon-DM interaction, and DM annihilation and decay modifies the previously mentioned allowed zone significantly. The modified regions in the $m_{\chi}$-$\sigma_{41}$ plane are described by the regions lying between the pairs of dashed lines of the same colour for individual cases. The area bounded by red dashed line corresponds to the allowed zone for the VDE Model I with $\eta=0.3$. The region between the blue lines represents the Model II for $\eta=0.3$ and $\omega_{\rm de}=-0.5$. Similarly, the green lines correspond to the Model II corresponding to phantom dark energy, {\it i.e.}, for $\omega_{\rm de}=-1.2$. The orange lines are for Model III with $\eta=0.2$ and $\Omega_{k,0}$. In all the four cases the PBH mass $M_{\rm BH}=10^{15}$ g, $\beta_{\rm BH}=10^{-29}$ and $\tau_{\chi}=10^{28}$s are chosen, while the value of $\langle \sigma v \rangle$ at different points are estimated for corresponding dark matter masses $m_{\chi}$. From this figure (\autoref{fig:mchi_sigma}) it can be seen that in absence of viscosity in dark energy, Hawking radiation from PBHs and  dark matter interactions and decay, the maximum possible value of dark matter particle mass is $\sim 3$ GeV. This agrees with several recent works \cite{rennan_3GeV,21cm_feb}. However, as the effect of viscosity in dark energy is considered along with the heating due to Hawking radiation, DM annihilation and decay, the maximum possible value of $m_{\chi}$ increases with increasing $\sigma_{41}$, and there is no maximum value of $m_{\chi}$. In addition, the allowed regions are slightly shifted toward the right side (higher value of $\sigma_{41}$) in viscous dark energy models compared to those for $\Lambda$CDM.
	
	\section{\label{sec:conc} Conclusions}
	
	In the present work the effects of viscous dark energy are investigated in the context of the global 21-cm signal.  We perform our
	analysis for three viscous dark energy model and evaluate their impact
	on the Hubble evolution and the 21-cm brightness temperature due to
	entropy production as a result of viscous flow.  We find that the Hubble parameter decreases with increasing viscosity parameter $\eta$ for lower values of $\eta$ at any fixed redshift $z$ in the case of Model I. But after a certain value of $\eta$ the Hubble parameter starts increasing rapidly. The minimum values of $H$ for individual redshifts trace a curve which signifies the departure from the $\Lambda$CDM behaviour at various redshift values. However, this curve smoothly approaches the $\Lambda$CDM value in the limit of vanishing viscosity. Similar characteristics are observed in the case of Model II and Model III as well, where it is again seen that higher values of viscosity could lead to significant departure of the Hubble parameter for various redshifts from the corresponding $\Lambda$CDM values.   
	
	The viscous flow of dark energy alters the Hubble evolution along with injecting additional entropy into the dynamics. It is observed that the net effect of viscosity cools down the baryons faster than for the case of the $\Lambda$CDM model, leading to significant modification of the 21-cm brightness temperature. Besides the impact of viscosity, the effect of baryon cooling due to baryon-dark matter interaction, and heating as an outcome of Hawking radiation from PBHs, DM annihilation and DM decay are also incorporated in our analysis. We perform a comparative study of the effect of viscosity with the effects due to the baryon-DM scattering, DM annihilation and decay, and PBH evaporation. We find that $T_{21}^{z=17.2}$ diminishes more at higher values of $\omega_{\rm de}$, while in the case of phantom model ($\omega_{\rm  de}<-1$) the variation of the brightness temperature with $\eta$ is comparatively small (for VDE Model II). It is also observed that the effect of curvature in the viscous dark energy model (Model III) slightly increases the values of $T_{21}^{z=17.2}$, while at a fixed value of curvature, the rate of change of $T_{21}^{z=17.2}$ with $\eta$ is considerable. 
	
	The dark matter profile plays an interesting role in the obtained value of $T_{21}^{z=17.2}$. We find that for higher values of mass $m_{\chi}$ ($\gtrapprox 1.5$ GeV) of the dark matter particles, a comparatively higher brightness temperature is obtained in presence of viscosity. However, for lower DM mass, higher values of viscosity parameter $\eta$ lie outside the permissible region according to the EDGES result. The interplay of the DM mass $m_{\chi}$ and the baryon-DM scattering cross-section $\sigma_{41}$ vis-a-vis the viscosity model parameters are presented in terms of a contour representation of the allowed region in $m_{\chi}-\sigma_{41}$ space. We find that the allowed zone moves toward higher values of $\sigma_{41}$ in presence of viscosity. For lower values of $\sigma_{41}$, both the upper and lower bounds of $m_{\chi}$ are less than that for the $\Lambda$CDM model (absence of viscosity in dark energy, PBH evaporation, DM annihilation and decay), while for higher values of $\sigma_{41}$, the upper bound of $m_{\chi}$ in VDE models becomes larger than 3 GeV which is the maximum allowed value of $m_{\chi}$ for the $\Lambda$CDM model.
	
	To summarize, in presence of viscosity in the dark energy fluid, the evolution of the Hubble parameter could get modified significantly leading to faster cooling of baryonic fluid. On the other hand, viscosity also produces some amount of entropy which essentially heats up both the baryon and dark matter fluids. Incorporating all of these effects along with the baryon heating due to PBH evaporation, and dark matter annihilation and decay, the resultant impact on the global signature of 21-cm absorption spectra is estimated. We also show that viscosity by itself cannot provide sufficient cooling to sustain the EDGES consequence ({\it i.e.} $-300\geq T_{21}\geq -1000$ at $z=17.2$). However, the net cooling due to the combination of VDE and that due to the baryon-DM interaction is capable of explaining the EDGES observation.	Our work should motivate further studies of viscous dark energy models by taking into account additional phenomena such as PBH accretion \cite{BH_21cm_0,Yang:2021agk}, and the spinning effect of PBHs \cite{saras_3_2,Cang:2021owu} in estimation of the 21-cm brightness temperature. Future space-based and ground-based experiments related to 21-cm physics will help to shed more light on such unknown aspects of cosmological dynamics.

	\section*{Acknowledgements}
	SSP thanks the Council of Scientific and Industrial Research (CSIR), Govt. of India, for funding through CSIR JRF-NET fellowship.
	

	\bibliographystyle{JHEP}
	\bibliography{PUB21}
	
\end{document}